\documentclass[acmtog, authorversion]{acmart}
\usepackage{booktabs} 
\usepackage{pifont}
\usepackage{amsmath}
\usepackage[export]{adjustbox}
\usepackage{array}
\newcolumntype{C}[1]{>{\centering\arraybackslash}m{#1}}

\begin{document}
	\title{Superimposition-guided Facial Reconstruction from Skull}
	
	\author{Celong Liu}
	\affiliation{%
		\institution{Louisiana State University}
		\city{Baton Rouge}
		\state{Louisiana}
		\postcode{70803}
	}
	\email{cliu32@lsu.edu}

	\author{Xin Li$^*$}
	\affiliation{%
		\institution{Louisiana State University}
		\city{Baton Rouge}
		\state{Louisiana}
		\postcode{70803}
	}
	\email{xinli@lsu.edu}

	\begin{abstract}
		\textbf{Abstract.} We develop a new algorithm to perform facial reconstruction from a given skull. This technique has forensic application in helping the identification of skeletal remains when other information is unavailable. Unlike most existing strategies that directly reconstruct the face from the skull, we utilize a database of portrait photos to create many face candidates, then perform a superimposition to get a well matched face, and then revise it according to the superimposition. To support this pipeline, we build an effective autoencoder for image-based facial reconstruction, and a generative model for constrained face inpainting. Our experiments have demonstrated that the proposed pipeline is stable and accurate. 
	\end{abstract}

	
	\maketitle
	
	\section{Introduction}
	
	Facial reconstruction from skull is a powerful tool to help forensic investigators identify skeletal remains when other information is not available. It has been successfully applied in many real forensic cases. Although in the past two decades, quite a few computer graphics based facial reconstruction algorithms have been developed, no existing method has reported adequate accuracy for law enforcement~\cite{Face:WilkinsonJA10}. In current forensic cases, facial reconstruction is still performed manually. 
	The commonly adopted reconstruction pipeline consists of three steps~\cite{Face:TaylorFaceBook05}: On the subject skull that needs identification, first, place landmarks on a set of anthropometric points; then, extend these landmarks following certain statistically standard tissue thickness; finally, produce a clay face model following these \emph{extended landmarks}. This generated face is the reconstructed face on this skull.  	
	
	\textbf{CG-based Facial Reconstruction.}
	Several systems have been developed to digitally mimic this procedure based on computer graphics and modeling techniques. A most commonly adopted strategy is to deform a template face surface~\cite{Face:Muller05ISPA} or tissue volume~\cite{Face:KahlerSIG03,Face:Jones01VMVC} to fit with the subject skull. 
	The deformation is governed by penalizing certain geometric smoothness energy so as to minimize the stretching of the transformation, while enforcing positions of extended landmark points calculated from tissue depths. 
	The limited constraint from the tissue depth information and smoothness criterion often makes face synthesis ill-posed and unstable. Hence, substantial refinement from modelers are often required to make these reconstructed faces realistic.
	
	\textbf{Learning-based Facial Reconstruction.}
	Another category of reconstruction algorithms are built upon statistical models of faces~\cite{Face:Claes06FSI,Face:PaysanDAGM09,cao2014facewarehouse}.

	In these approaches, each face is abstracted using a high dimensional vector composed of 3D coordinates of (feature) points, and the database can be modeled using principal component analysis. A new face is then defined as linear combinations of the principal components. However, these reconstructions often produce a globally averaged geometry with characteristic details smoothed out. 
	Unfortunately, such an ``average face'' is not very useful, as facial characteristics are critical for recognition. 
	
	\textbf{Ambiguity in growing face from a skull.} 
	A major limitation of existing facial reconstruction algorithms, which directly reconstruct the face from a skull, is the ill-poseness of face synthesis from limited constraint of tissue depth information. 
	Although the general geometry of the face can be mostly determined by the skull, certain feature regions, such as lip shapes and eye brows, cannot be inferred from the bones. Therefore, their modeling often needs to rely on artistic interpretation of the forensic specialists, which is not only subjective (hence, reconstructions done by different modelers could be different), but also difficult to rigorously formulate.
	
	\textbf{Our idea.}
	To overcome this problem, we propose to solve the facial reconstruction through a different approach. Instead of directly reconstructing the face from a skull, we develop a novel three-step reconstruction pipeline. First, we do an \emph{image-based facial reconstruction} to generate many face candidates from a database of images. Second, we perform a \emph{skull-face superimposition} to compute the likelihood of each reconstructed face matching with the given querying skull. Finally, from a well (best) matched face, we do a \emph{face re-synthesis} to revise the face geometry according to the superimposition result. This new approach can effectively reduce the ambiguity of direct face reconstruction from limited set of constraints. 
	
	The \textbf{main contribution} of this work includes: (1) a new facial reconstruction pipeline to produce a realistic face according to the given skull; (2) a restricted generative model to support geometry-guided face inpainting; (3) an integrated face autoencoder for effective and stable image-based facial reconstruction.
	
	The remainder of this paper is organized as follows. We review related work in Section~\ref{Sec:RelatedWork}, then elaborate our image-based reconstruction algorithm in Section~\ref{Sec:Reconstruction}, and the skull-face superimposition and skull-guided face re-synthesis in Section~\ref{Sec:FaceInpainting}. We report the experimental results in Section~\ref{Sec:Results} and conclude the paper in Section~\ref{Sec:Conclusions}.
	
	\section{Related Work}
	\label{Sec:RelatedWork}

	Our facial reconstruction pipeline mainly involves developing two technical components: 3D face reconstruction from an image, and face inpainting. We review recent related work in these two topics. 
	
	\subsection{Face Reconstruction}	
	Classic image-based face reconstruction is based on various morphable models~\cite{blanz1999morphable,cao2014facewarehouse,paysan20093d}, where a parametric template model is deformed to fit a given image. Most classic morphable models are based on PCA, which unfortunately, has limited capability of describing face details (that are critical for recognition).  
	A recent parametric face model, FLAME~\cite{FLAME:SiggraphAsia2017}, decomposes the face into shape, pose, and expression parameters (blendshapes). This model is much more expressive than PCA, and hence, can provide more realistic and accurate face description. 
	However, a limitation of all these model fitting based approaches is their sensitivity to template selection. When the image and selected template are not similar, the model fitting often converges to a local optimum and may produce a 3D face that does not match well with the image.
	
	Another category of approaches is deep learning based methods. Compared with morphable models, learning-based methods have two general advantages. (1) It is often more stable, due to its less sensitivity to the initialization of model parameters (i.e., selection of template). (2) It is more efficient, because after training, its parameter estimation is much faster than morphabale models which require iterative optimizations. Hence, multiple face modeling systems have been built through deep learning, using, for example, multi-task Convolutional Neural Network (CNN)~\cite{ranjan2017all}, CNN cascades~\cite{wang2014facial, guo2018cnn}, Restricted Boltzmann Machines~\cite{wu2015discriminative}, and recurrent network with long-short term memory (RNN-LSTM)~\cite{zhao2018robust}. But these networks need to be trained in a supervised manner, and are currently used to only produce sparse information (features) of the face. 	
	In our problem, we need a dense reconstruction that produces the full face model. 
	Due to the limited availability of large volume of high-resolution 3D face scans  (whose acquisition is much more expensive than that of 2D portrait photos), an unsupervised approach is more desirable. 
	
	Unsupervised face reconstruction can be achieved by using a geometric Auto-Encoder (AE). Two recent notable frameworks are the fully CNN-based autoencoder (CAE) (e.g. the Deep Face Encoder~\cite{gao2015single}) and the Model-based autoencoder (MAE) (e.g. Model-based Face Autoencoder MOFA~\cite{tewari2017mofa}). The CAE uses CNN for both encoding and decoding, while the MAE uses CNN only for encoding and uses a parametric face model for deconding/reconstruction. 
	CAE usually cannot guarantee the semantic meaning of the code layer parameters; and they need to train enormous sets of unintuitive CNN weights. In contrast, MAE can avoid such disadvantages, because they already integrate some prior knowledge of human faces. Hence, MAE does not need that big amount of data for training and currently produces the state-of-the-art reconstruction results. 
	
	\subsection{Face Re-synthesis}
	\label{Sec:FaceResynthesis}
	
	The face re-synthesis problem we aim to solve here is to revise specific face regions following geometric constraints from the skull. This problem can be considered as a face inpainting problem, which first removes unmatched regions, then re-generate them under certain geometric constraints.  
	
	A direct method to inpaint a 3D face is to fit a statistical or parameterized 3D face model~\cite{blanz1999morphable,cao2014facewarehouse,FLAME:SiggraphAsia2017} onto the corrupted face. The model parameters should be estimated from remaining face points together with the extra skull constraints. 
	However, the parametric space and the space of realistic faces are often not bijectively mapped. A set of model parameters computed by fitting existing face points and extra constraints may not map to a realistic face, and we could end up getting faces with significant artifacts. 
	
	Recently, deep learning based techniques have demonstrated great success in image and geometric inpainting. 
	For 3D inpainting, generative models~\cite{wu2016learning, smith2017improved} have been developed, and they use voxelized objects from database to train deep neural networks. 
	However, these voxelized models do not provide enough details in describing fine characteristics of human faces. Furthermore, a huge amount of 3D face scanning is needed to built an effective 3D face inpainting system, but such a dataset is currently not publicly available. 
	Therefore, direct 3D face inpainting has not reached the same accuracy level of 2D face image inpainting. 
	In this work, we convert the inpainting problem to 2D, and utilize the state-of-the-art 2D image inpainting techniques to do the face synthesis, then reconstruct the facial geometry in 3D. 
	
	Deep learning based image inpainting techniques can be classified into non-generative and generative approaches. Non-generative approaches, such as \cite{mairal2008sparse,xie2012image,ren2015shepard,liu2016learning}, usually infer the unknown region by finding, copying, then refining a local patch with a similar structure from a model learned from a database. 		 These local-patch based strategies work better for holes that are small or have simpler local structural patterns. But they may not work well in repairing big/complex holes or corrupted faces which possess both locally and globally complex characteristics.
	
	Generative model based inpainting currently produces the-state-of-art results in face image inpainting. 
	The basic idea is to train a deep generative model (using e.g. Generative Adversarial Networks (GANs)~\cite{goodfellow2014generative, radford2015unsupervised}), and construct a latent space and a generator $G$. Then, map the corrupted image to its nearest point $z$ in the latent space, and use $G(z)$ to produce a globally realistic inpainted face image. DCGAN~\cite{radford2015unsupervised} is shown to be effective in building good face image generators. 
	The Context Encoder (CE)~\cite{pathak2016context} uses such a GAN to build the context generator, and it maps a corrupted image to its corresponding latent variable that has smallest context difference in the given image's non-missing region.
	However, since the latent space is usually a bigger (higher-dimensional) space than the space of realistic face images, an arbitrary latent variable $z$ may not always corresponds to a realistic image.  
	Therefore, images generated by the GAN could still be unrealistic (e.g. blurry).
	More recently, Yeh et al.~\cite{yeh2017semantic} introduce a prior loss when finding the latent variable of a corrupted image. 
	Ensuring a small prior loss (small loss from the discriminator of the GAN) makes the latent variable $z$ to not only have small context loss, but also generate a realistic face.  
	This greatly improves the authenticity of the inpainted context and produces the state-of-the-art face inpainting results.
	However, this inpainter only encodes 2D context from non-missing regions and hence, only provides us an arbitrarily repaired face, we need to modify it by further incorporating extra 3D geometric constraints, to get the face that aligns with the given skull. 
	
\section{Image-based Facial Reconstruction}
\label{Sec:Reconstruction}
	
\begin{figure}[h]
		\centering
		\includegraphics[width=0.5\textwidth] {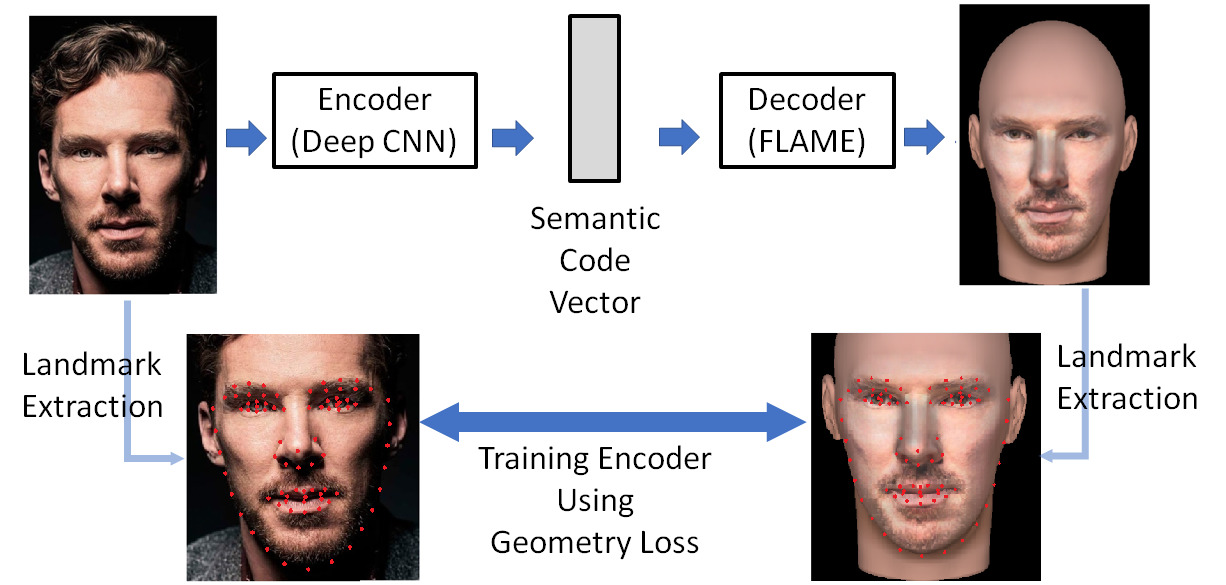}
		\caption{The Pipeline of Our Proposed Auto-encoder.~\label{Fig:AutoEncoderPipe}}
	\end{figure}

Inspired by the structure introduced in Model-based Face Autoencoder (MOFA)~\cite{tewari2017mofa}, we build a deep auto-encoder for facial reconstruction. 
In existing autoencoders~\cite{tewari2017mofa,gao2015single}, faces are often represented using PCA, and also, a big portion of the parameters are used in describing illumination and skin reflection. 
In our problem, our focus is to get accurate facial geometry. Hence, we modify the design of this autoencoder by (1) using a geometry-based loss function, and (2) adopting a more accurate face parametric model. Our design is illustrated in Fig.~\ref{Fig:AutoEncoderPipe}. 
The encoder converts a face image into a semantic vector; and the decoder generates a 3D face, then re-synthesizes an image of this face. 
The synthesized image is compared with the input image, using a geometric loss, to refine this autoencoder.

\subsection{Building Database by Data Fusion}
\label{Subsec:BuildingDatabase}

\textbf{3D face dataset}. To train a fine parametric model, we need a big volume of 3D registered head/face data. Most learning-based algorithms are built on their own datasets, but many of these datasets are not publicly available. We integrated faces from 7 relatively big public datasets: BU-3DFE~\cite{yin20063d}, BFM~\cite{paysan20093d}, FRGC~\cite{phillips2005overview}, Magna~\cite{evison2008magna}, Texas 3DFRD~\cite{gupta2010anthropometric}, BOSPHORUS~\cite{savran2008bosphorus} and D3DFACS~\cite{FLAME:SiggraphAsia2017}. Among them, D3DFACS are complete head scans, which are best suitable to match with skulls, while others are frontal face scans. Therefore, we use D3DFACS as the template datasets, and consistently register and parameterize all other face data onto the head geometry in D3DFACS. We implement a dense correspondence algorithm~\cite{gilani2017dense} to register all the 3D faces. By integrating all these $7$ datasets we have obtained a database of $30k$ 3D faces (or more rigorously speaking, 3D heads), parameterized according to D3DFACS.
	
\textbf{2D face images}. 
To train the face auto-encoder, we also need an image corpus. We combine four datasets: CelebA~\cite{liu2015deep}, LFW~\cite{huang2007labeled}, Facewarehouse~\cite{cao2014facewarehouse}, and 300-VW~\cite{shen2015first}. 
We detect the faces on all these images using the Haar Cascade Face Detector~\cite{viola2001rapid}, crop the background, then normalize each image to $240\times 240$ pixels. Then, by applying the facial landmark detector~\cite{saragih2011deformable}, we get a consistent annotation on all face images automatically. 
In total, this provides us $147k$ images, which are randomly partitioned for training ($142k$) and evaluation ($5k$) in our experiments.

\subsection{Semantic Code Vector and Encoder}
The face encoder extracts features from the input face image to compose a \emph{semantic code vector}. This vector contains two types of information: 
(1) facial geometry: a set of parameters that can be used to reconstruct the 3D face, and   
(2) rendering information: parameters such as the camera poses and scene illumination. 

We adopt FLAME~\cite{FLAME:SiggraphAsia2017}, a state-of-the-art parametric face model, to describe the facial geometry. 
In FLAME, a face geometry is described by a function $M(\alpha, \delta, \theta): \mathcal{R}^{|\alpha|\times|\delta|\times|\theta|} \rightarrow \mathcal{R}^{3N}$, where $\alpha, \delta$, and $\theta$ are the coefficients describing face shape, expression, and pose, respectively. $\theta \in \mathcal{R}^{3K+3}$ indicates $K+1$ rotation vectors, where $K$ is the number of joints (each rotation is a $3$-dimensional vector, plus one global rotation).
The rendering information is parameterized by a camera rotation $T \in SO(3) \in \mathcal{R}^3$, a camera translation $t \in \mathcal{R}^3$, and the scene illumination coefficients $\gamma \in \mathcal{R}^{27}$. 
	
We need to choose an appropriate dimension size for each parameter component. The size of $T$, $t$ and $\gamma$ are fixed. Following the experiments in \cite{FLAME:SiggraphAsia2017}, $\theta=15 \; (K=4)$ is sufficient,
and choosing $|\alpha|=|\delta|=90$ can guarantee that $99.9\%$ of the fitting errors are less than $1mm$, which is smaller than our tissue depth error threshold (Section~\ref{Sec:Superimposition}). Hence, we set them using these values, and the final dimensionality of our semantic code vector $x = (\alpha, \delta, \theta, T, t, \gamma)$ is $228$. Our face meshes are consistently sampled using $N=94,154$ vertices. Therefore, our face model is a function $M(\alpha, \delta, \theta): \mathcal{R}^{195} \to \mathcal{R}^{N}$. 

Using a CNN we can build the encoder to extract the semantic code vector from a face image.	According to \cite{tewari2017mofa}, VGG-Face~\cite{Parkhi15} gives the best face recognition result among various CNN structures. 
We also adopt VGG-Face as our architecture, but modify its fully connection layer and change the output to $228$ dimension. 
	
\subsection{Decoder}
\label{Subsec:Decoder}

Taking the semantic vector $x$ as the input, our decoder first generates a 3D face using the FLAME coefficients ($\{\alpha, \delta, \theta\}$), then use the rendering parameters ($\{T, t, \gamma\}$) to synthesize an image of this face. 
The calculation of image synthesis is fully analytic and differentiable, as is derived in the following. 

\begin{enumerate}
	\item \textit{Perspective Camera}. A pinhole camera model follows a perspective projection $\Pi: \mathcal{R}^3 \rightarrow \mathcal{R}^2$ to map from the camera space (camera coordinates) to screen space (image coordinates). The position and orientation of the camera in the world coordinates is given by a rigid transformation, the rotation $T \in SO(3) \in \mathcal{R}^3$ and global translation $t \in \mathcal{R}^3$, by $\Phi_{T,t}(p) = T^{-1}(p-t)$ for any point $p$ in the world coordinates. Finally, $\Pi \circ \Phi_{T,t}(p)$ maps $p$ to its image coordinates.
	\item \textit{Illumination}. The illumination model is a
	Spherical Harmonics model. Here, we assume distant low-frequency illumination and a purely Lambertian surface reflectance. Thus, we evaluate the radiosity at vertex $v_i$ with surface normal $n_i$ by $$C(n_i, \gamma) = r\cdot \sum_{b=1}^{B^2}\gamma_b H_b(n_i),$$
	where $H_b: \mathcal{R}^3 \rightarrow \mathcal{R}$ are the SH basis functions, $\gamma_b \in \mathcal{R}^3$ ($B=3$ bands) are
	coefficients that parameterize colored illumination using the red, green, and blue channel. $r$ is the face reflectance. Instead of predicting $r$ for each vertex, we use a fixed face reflectance following the implementation of original 3DMM~\cite{blanz1999morphable}.
\end{enumerate}

\textbf{Image Formation}. We render a face image using the aforementioned camera and illumination model. Hence, in the 
forward pass $\mathcal{F}$, we compute the screen space position $u_i(x)$ and associated pixel color $c_i(x)$ for each vertex $v_i$:
$$\mathcal{F}_i(x) = [u_i(x), c_i(x)]^T \in \mathcal{R}^2,$$
$$u_i(x) = \Pi \circ \Phi_{T,t}(M(\alpha, \delta, \theta)),$$
$$c_i(x) = C(Tn_i(\alpha, \delta, \theta), \gamma).$$
Here, $Tn_i$ transforms the world space normals to camera space and $\gamma$ models illumination in camera space.

We can implement a backward pass that inverts the image formation:
$$\mathcal{B}_i(x) = \frac{d\mathcal{F}_i(x)}{d(\alpha, \delta, \theta, T, t, \gamma)}.$$
This computes gradients of the image formation model with respect to parameters in the semantic code vector. 
	
\subsection{Geometry-based Loss}
When building the encoder in MOFA~\cite{tewari2017mofa}, the loss is calculated using a pixel-to-pixel color difference. A limitation of this loss is that it could cost the majority of parameters in the encoder's network being used to model the rendering (camera, illumination, and skin reflection). And this may affect both efficiency and effectiveness in geometry reconstruction. Hence, we employee a sparser geometric loss function based on detected facial landmarks~\cite{saragih2011deformable}: Firstly, $66$ typical face landmarks are extracted; then among them, after merging too-closed pairs, a subset of $46$ landmarks are preserved~\cite{saragih2011deformable}. 
	
	\begin{figure}[!htb]
		\centering
		\begin{tabular}{cccc}
			\includegraphics[height=0.105\textwidth] {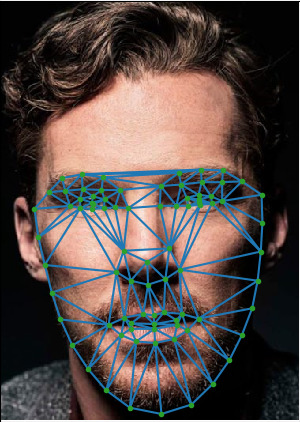} &
			\includegraphics[height=0.105\textwidth] {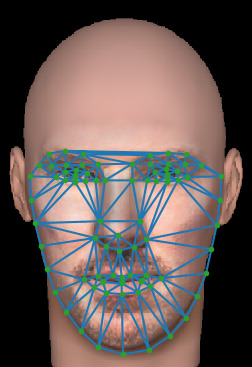} &
			\includegraphics[height=0.105\textwidth] {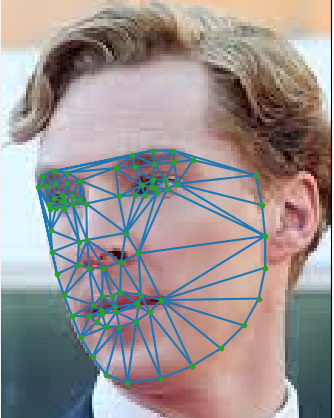} &
			\includegraphics[height=0.105\textwidth] {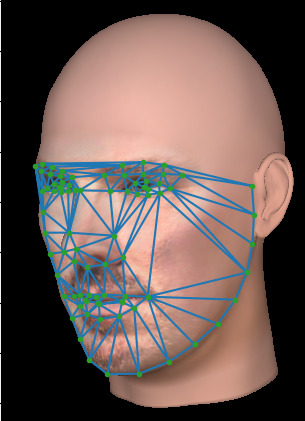} \\
			(a) & (b) & (c) & (d)\\
		\end{tabular}
		\caption{Landmarks are consistently detected on both original portrait photos (a,c) and their corresponding synthesized images (b,d). Note that when training the autoencoder, the synthesized image is rendered from the same camera angle, and has a same landmark distribution with the original photo.~\label{Fig:Landmark}}
	\end{figure}
	
	From both the input image, we extract these landmarks $P$ and from the synthesized image we extract the corresponding landmarks $P'$. We compute a Delaunay triangulation $C$ on $P$, then transfer the connectivity to $P'$, denoted as $C'$, respectively. 
	Fig.~\ref{Fig:Landmark} illustrates these landmarks and their triangulations.
	Finally, the loss is defined on the edge set $C=\{ e \}$ and $C'=\{ e' \}$, by 
	\begin{equation}
	E_{loss}(x) = w_m E_m(x) + w_r E_r(x),
	\label{Eqn:E_Loss}
	\end{equation}	
	where $E_m(x) = \sum_{i=1}^{|C|}(|e_i| - |e_i'|)^2$ is the geometry term measuring the change of edge lengths, and $E_r(x) = \sum_{k=1}^{90}\alpha_k^2 + \sum_{k=1}^{90}\delta^2 + \sum_{k=1}^{15}\theta^2$ is the regularization term.
	
	\textbf{Backpropagation.} To enable training based on stochastic gradient descent, during backpropagation, the gradient of $E_{loss}$ (Eq.~(\ref{Eqn:E_Loss})) is passed backward to our model-based decoder and is combined with $\mathcal{B}_i(x)$ using the chain rule.
		
	\subsection{Reconstruction Refinement using Multiple Images}
	
	While this auto-encoder can reconstruct 3D face from any given face image with good details, when the photo is not properly taken (with important characteristics missing), the reconstruction will inevitably be less accurate. To support an iterative refinement when needed and a generally more stable reconstruction, we also design the auto-encoder to take in multiple images. 
	The previously formulated encoder converts a face image into the geometric parameters
	$G = \{\alpha, \delta, \theta\}$ and rendering information $R = \{T, t, \gamma\}$. We denote this original loss function as $E_{loss}(G, R)$.
	
	\textbf{Training the New Multi-image Encoder.} Suppose multiple images $\{I_j, j\in[1, m]\}$ are available for one person, we can extract multiple semantic vectors
	$x_1=\{G_1, R_1\}, x_2=\{G_2, R_2\}, \ldots, x_m=\{G_m, R_m\}$.
	The face geometry should be as close as possible $G_1=G_2\ldots=G_m$. Hence, we define the following loss on all the $x_j$, 
	\begin{equation}
	E_{loss}^{mult}(G; R_1, \ldots, R_m) = \sum_{j=1}^{m}E_{loss}(G, R_j),
	\label{Eqn:E_Loss_Mult}
	\end{equation}
	which uses only one geometry parameter $G$ to model faces from all these images.	
	And with this $E_{loss}^{mult}$, we change the training into two stages.	
	\begin{itemize}
		\item[1)] \textbf{Stage 1.} Train the original auto-encoder discussed in the previous section, save all the semantic vectors as initial values for the next stage.
		\item[2)] \textbf{Stage 2.} Group the training images by person. For images from a same person $p$,  $\{I_1^p, I_2^p, \ldots, I_{n_p}^p\}$, use the new loss function 
		$E_{loss}^{multi}$ to refine the network, enforcing the reconstructed geometry for the same person to be the same.
	\end{itemize}
	
	\textbf{Decoding.} To use this new encoder to reconstruct a face:
	\begin{enumerate}
		\item Feed the encoder with a first face image $I$, get the semantic vector, which can be used to reconstruct the 3D face mesh;
		\item if more images are available, first get multiple semantic vectors, which may result in different geometries, then solve the optimization in Eq.~(\ref{Eqn:E_Loss_Mult}) using rendering parameters $\{R_i\}$, and return the shared geometry parameter 
		$\hat{G} = \arg \min_{G} E_{loss}^{mult}$.
	\end{enumerate}
	
\begin{figure}[h]
		\centering
		\addtolength{\tabcolsep}{-4pt}
		\begin{tabular}{cccccc}
			\includegraphics[height=0.1\textwidth] {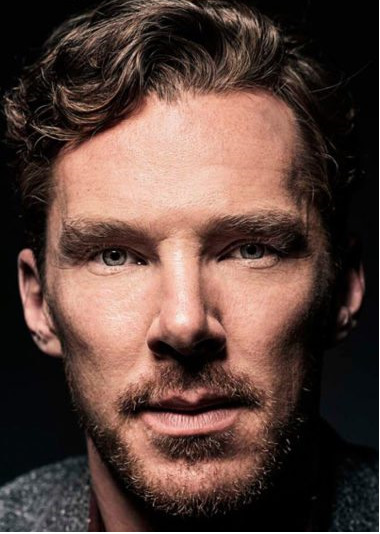} &
			\includegraphics[height=0.1\textwidth] {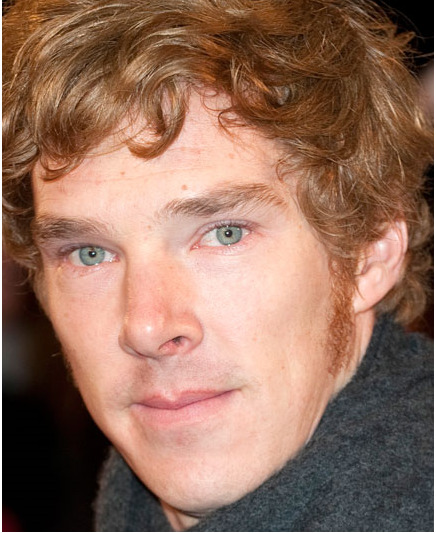} &
			\includegraphics[height=0.1\textwidth] {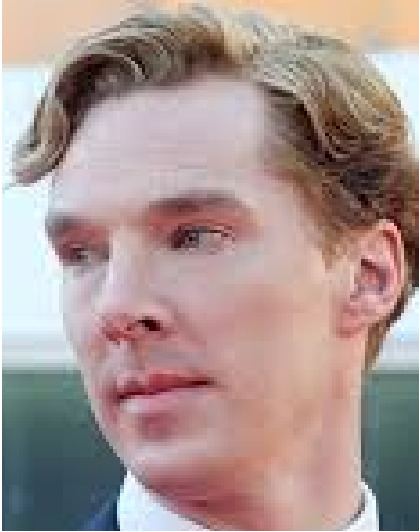} &
			\includegraphics[height=0.1\textwidth] {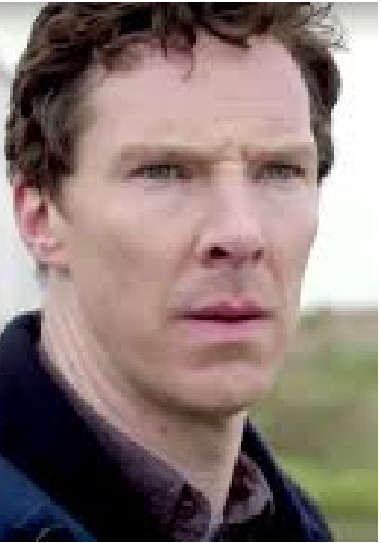} &
			\includegraphics[height=0.1\textwidth] {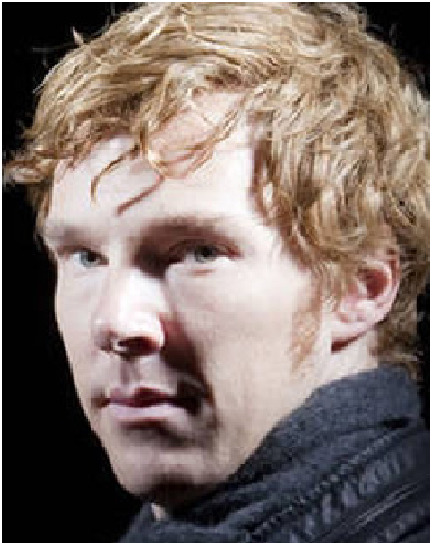} &
			\includegraphics[height=0.1\textwidth] {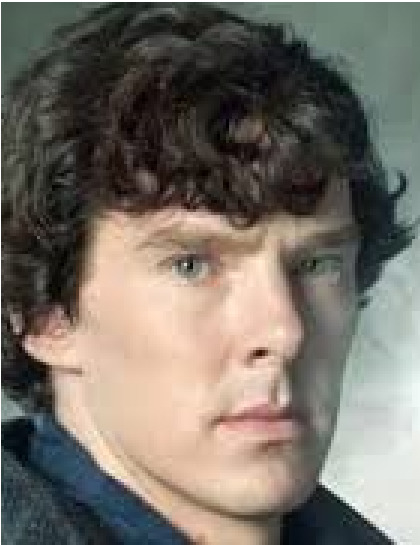} \\
			(a) & (b) & (c) & (d) & (e) & (f)\\
			\includegraphics[height=0.1\textwidth] {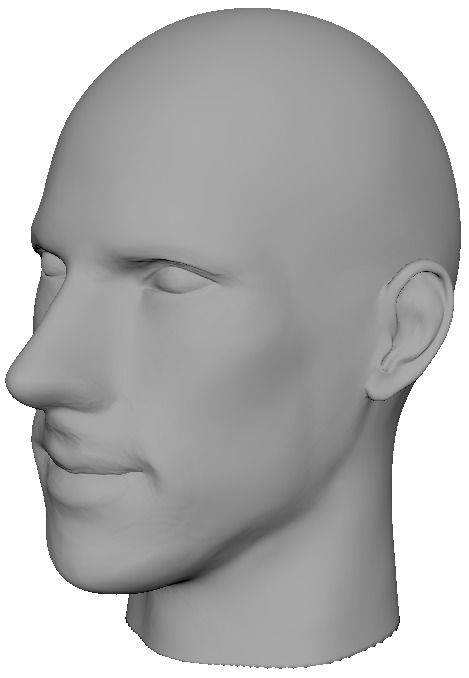} &
			\includegraphics[height=0.1\textwidth] {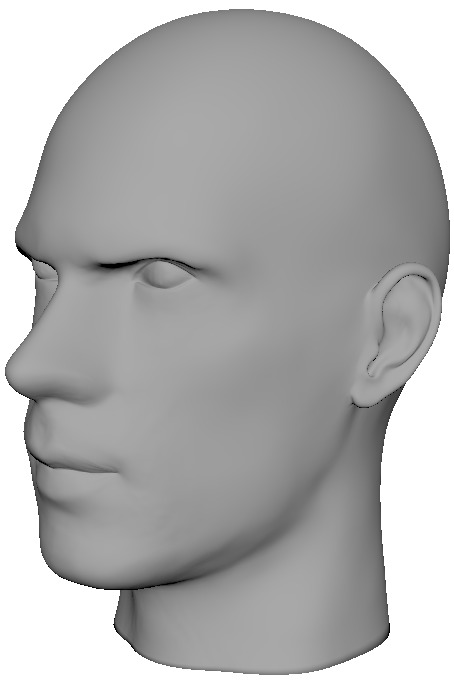} &
			\includegraphics[height=0.1\textwidth] {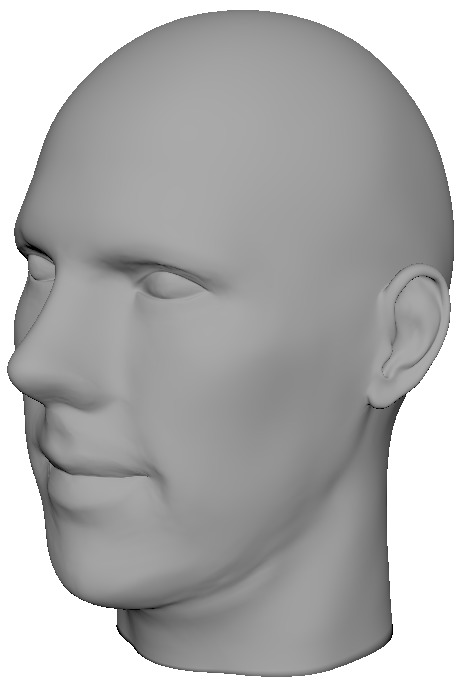} &
			\includegraphics[height=0.1\textwidth] {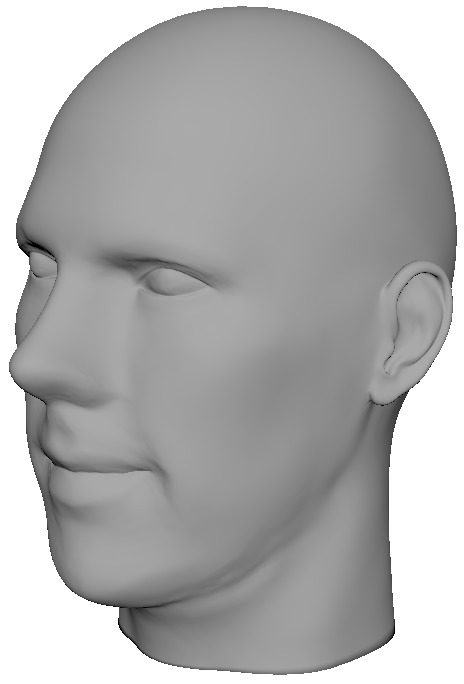} &
			\includegraphics[height=0.1\textwidth] {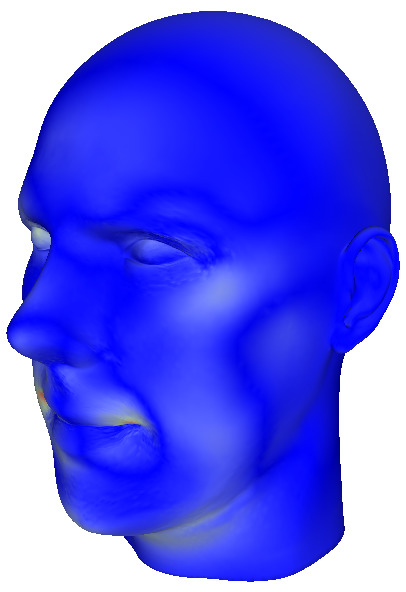} &
			\includegraphics[height=0.1\textwidth] {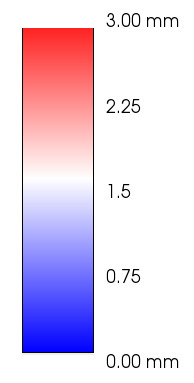} \\
			(g) & (h) & (i) & (j) & (k) & \\
		\end{tabular}
		\caption{Reconstructions of a same person's face using single and multiple images. (a)-(f) are 6 photos for one person; (g) is the reconstructed face just from (a); (h) is the reconstructed face just from (d); (i) is the reconstructed face from (a)-(c); (j) is the reconstructed face from (d)-(f); (k) color-encodes the point-to-point deviation between (i) and (j). The maximal deviation is $2.9mm$.~\label{Fig:MultImage}}
	\end{figure}

	Fig.~\ref{Fig:MultImage} demonstrates the effect of this multi-image refinement. Faces (g) and (h) are the reconstruction results from image (a) and image (d), respectively. They have relatively big difference. 	
	Face (i) is reconstructed from the three images (a-c), and Face (j) is reconstructed from images (d-f). 
	When multiple images are used, separately reconstructed faces converge stably into a similar geometry. 
	Although the two separate image sets (a-c) and (b-d) are randomly selected, the reconstructed face (i) and (j) is very similar. Their vertex-to-vertex deviation is calculated and color-encoded in (k). The maximal deviation is $2.9mm$ (which is significantly smaller than face reconstruction error, see Table~\ref{Tab:ReconComp}). This shows that this multiple-image model produces stable and converged reconstruction results. 
	
\subsection{Generating 3D Face Candidates}
\begin{figure}[h]
	\centering
	\begin{tabular}{cc}
		\includegraphics[height=0.18\textwidth] {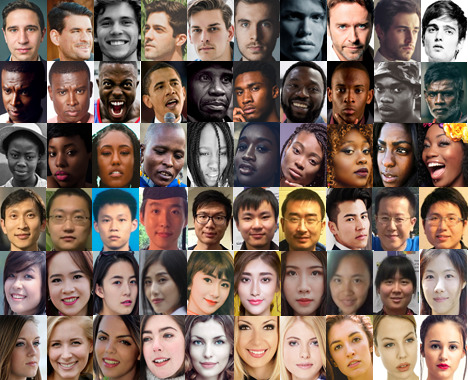} &
		\includegraphics[height=0.18\textwidth] {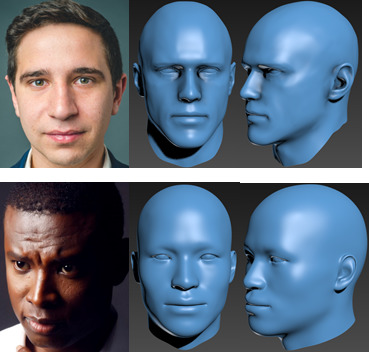} \\
	\end{tabular}
	\caption{Images downloaded from Internet and some reconstructed 3D faces.~\label{Fig:QueryingDatabase}}
\end{figure}

Given a database of portrait photos, using this auto-encoder, we can now reconstruct their 3D faces. These faces are used as potential face candidates to match the given querying skull, and suggest the skull's possible face appearance. The best matched ones will be used as starting faces to synthesize the final face in the next section. In our experiments, we have reconstructed $80k+$ 3D face candidates using portrait images downloaded from the Internet.  Fig.~\ref{Fig:QueryingDatabase} illustrates some reconstruction results. 
	
\section{Skull-guided Face Re-synthesis}
\label{Sec:FaceInpainting}
	
From the big amount of reconstructed face candidates, we match each of them with the given querying skull, through performing a skull-face superimposition (Section~\ref{Sec:Superimposition}). Usually, when the deceased is not in the database, no face would perfectly match the skull, we then pick a face and modify its poorly matched regions, to re-synthesize a new face according to the skull. We do this re-synthesis through an inpainting strategy following guidance from the superimposition (Sections~\ref{SubSec:FaceInpaintingIdea}-\ref{Subsec:Stability}).
	
\subsection{Face-Skull Superimposition}
\label{Sec:Superimposition}
	
\begin{figure}[h]
	\centering
	\begin{tabular}{ccc}
		\includegraphics[height=0.1\textheight] {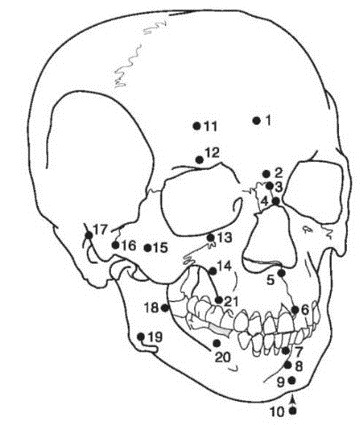} &
		\includegraphics[height=0.1\textheight] {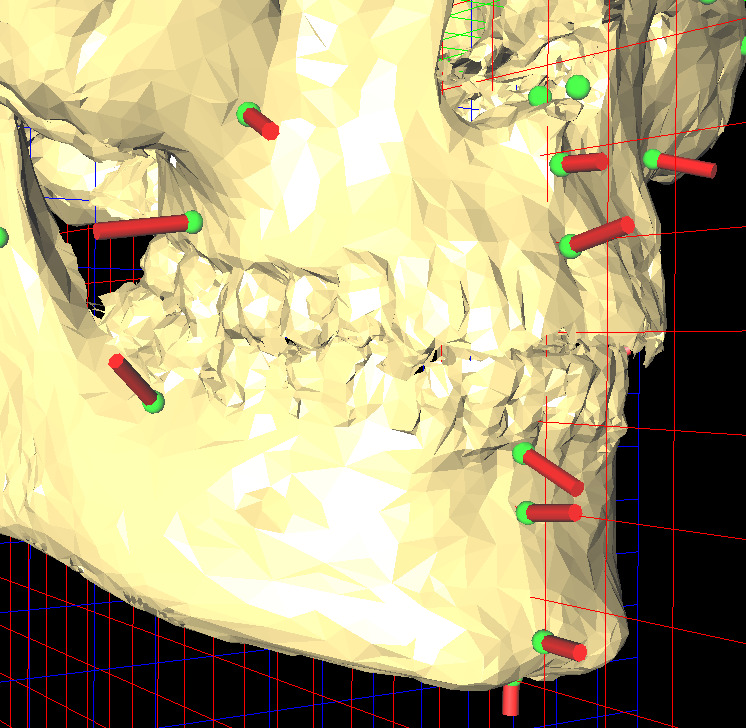} &
		\includegraphics[height=0.1\textheight]{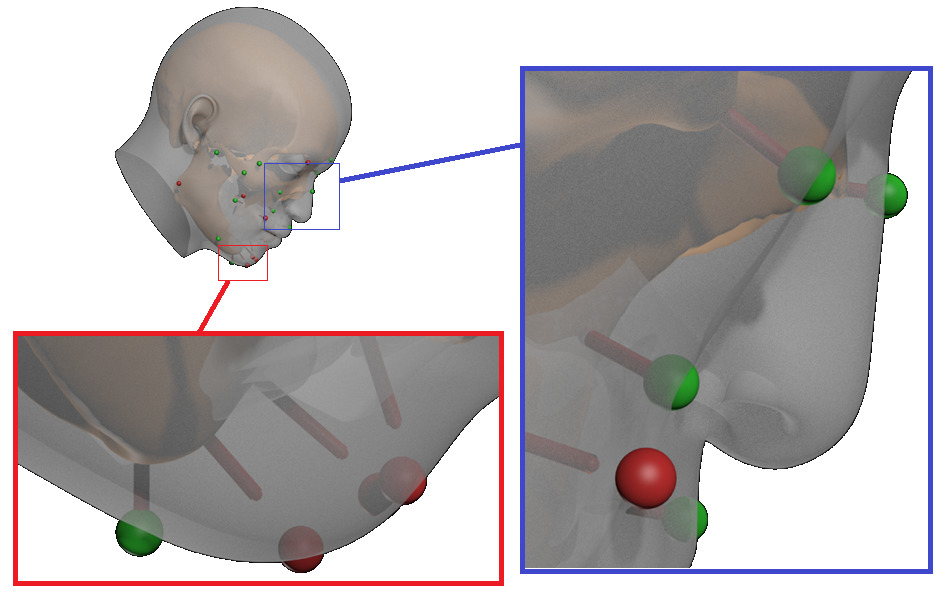} \\
		(a) & (b) & (c) \\
	\end{tabular}
	\caption{(a) Anthropometric landmarks on a skull; (b) tissue depths defined on landmarks; (c) matched (green) and unmatched (red) extended landmarks in the skull-face superimposition.~\label{Fig:Superimposition}}
\end{figure}

Following the commonly adopted tissue-based facial reconstruction procedure~\cite{Face:TaylorFaceBook05}, we consider a set of anthropometric landmarks on a skull~\cite[page ~350 ff.]{taylor2000forensic} (also see Fig.~\ref{Fig:Superimposition}(a)). Each landmark is associated with a vector along the skull surface normal direction, corresponding to the direction of thickness measurements (see Fig.~\ref{Fig:Superimposition}(b)). The statistically standard tissue thickness was measured on different landmarks and recorded (e.g. \cite{rhine1984forensic}). 
Such depths on landmarks can be used to directly perform facial reconstruction from the skull, as well as evaluate how well a face anatomically matches with a skull. 
	
To perform a face-skull superimposition, we model an outward vector on each landmark $m_i$ along the skull surface's normal direction with its length being the expected tissue thickness. Then we call such an end point an \emph{extended skull landmark} $n_i$. 
If the given skull and the reconstructed face match well with each other, then the extended skull landmark will be close to the face, or more specifically, to a corresponding anthropometric point $p_i$ on the face. Since all our face models are consistently parameterized, after manually labeling these anthropometric points $\{p_i\}$ on one template face, the annotation can automatically propagate to all the other faces. 
If the Euclidean distance $\|n_i p_i\| <$ a threshold $\eta_i$, we get a matched landmark (e.g., green points in Fig.~\ref{Fig:Superimposition}(c)), otherwise, we have an unmatched landmark (e.g., red points).
Following \cite{damas2011forensic}, the \emph{superimposition score}, or skull-face matching probability, can be estimated using the ratio of the green to red points, or $S = M / (M+U)$, where $M$ and $U$ are the numbers of matched and unmatched landmarks.

For a given skull, we can perform its superimposition with each face in the database, and report a list of best matched faces. They have higher probability to look like the deceased.  
	
In many cases, the skull's corresponding face is not included in the face image database. Then, even the best matched faces have their superimposition scores $S < 100\%$. We can revise the reconstructed face according to the superimposition. Our idea is to preserve the well matched region, and re-synthesize the unmatched part following the extended landmarks. We elaborate our strategy to accomplish this through face inpainting in the following section. 
	
\subsection{Face Re-synthesis through Inpainting}
\label{SubSec:FaceInpaintingIdea}

To revise the reconstructed face following the skull, we remove the unmatched regions on the face, then inpaint (re-fill) them following the extended landmarks grown from the skull. 
	
\textbf{Face Inpainting by 3D Model Fitting.} A direct strategy to fill the removed region is through model fitting (using a parametric face model such as FLAME). Without considering the guidance from the skull, the filled regions would not match the extended landmarks. We should perform a model fitting integrating  geometric constraints from the extended landmarks. 
However, as shown later in this section, a key \emph{limitation} of this constrained model fitting approach is that it could result in artifacts near constrained landmarks and hence, less realistic faces. 
	
\textbf{Generative Inpainting.} 
Recently, the Generative Adversarial Network (GAN) has demonstrated great success in image inpainting. 
By building a GAN face generator, we synthesize the missing region of a face following the constraints from the non-missing region.
Its main advantage over the constrained model fitting approach is that generative inpainting produces more realistic faces. 
As discussed in Section~\ref{Sec:RelatedWork}, due to the limited available data, direct 3D face generative inpainting has not be able to produce results as good as 2D face image inpainting. We therefore perform a two-step inpainting: first, do the inpainting on a face image, then, use our face reconstruction algorithm (Section~\ref{Sec:Reconstruction}) to obtain the 3D face. 
	
A state-of-the-art GAN-based face image inpainting algorithm is recently suggested by Yeh et al.~\cite{yeh2017semantic}. In ~\cite{yeh2017semantic}, a GAN is trained with face images, then the generator $G$ is used to reconstruct the face image after finding a latent variable $z$ by solving an optimization that minimizes the distance between a generated authentic image $G(z)$ and the corrupted image in the non-missing region.
We adopt this idea, but make a modification in building the training datasets. We train the GAN using a set of re-rendered, normalized face images, rather than using regular face photos. This could improve the performance of the GAN in generating 3D face models, rather than emphasizing the generation of realistic 2D face images.
	
\textbf{Generative Image Inpainting with Geometric Constraints.}
Besides generating a realistic face, in our problem, we also need to control the shape of the final face so that it passes through the extended landmarks. 
To ensure this geometric constraint, we design a superimposition error term to evaluate the deviation of the face from the extended landmarks. 
This new generative inpainting strategy allows us to generate a face that is both \emph{realistic} and \emph{having small superimposition error} with the skull. We elaborate our proposed inpainting strategy in the following sections.
	
\subsection{Face Segmentation} 
\begin{figure}[h]
		\centering
		\begin{tabular}{ccc}
			\includegraphics[height=0.15\textheight] {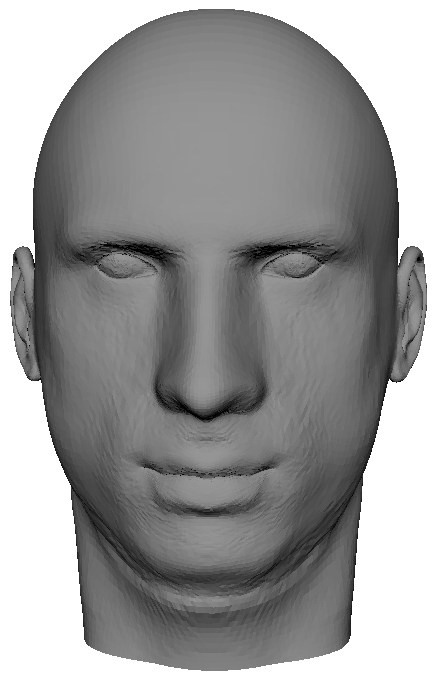} &
			\includegraphics[height=0.15\textheight] {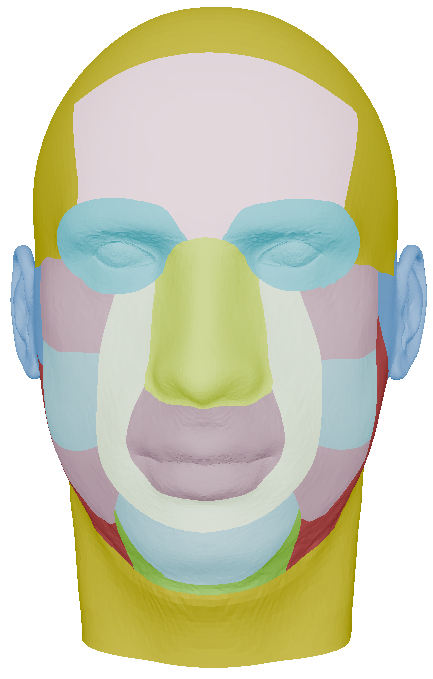} &
			\includegraphics[height=0.15\textheight] {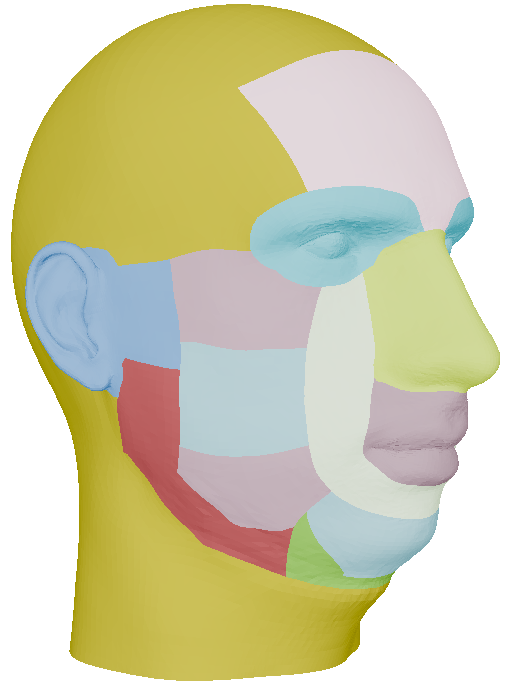} \\
			(a) & (b) & (c) \\
		\end{tabular}
		\caption{Segmenting the Face (Head) Model into Subregions.~\label{Fig:Segmentation}}
	\end{figure}

To revise the specific subparts of the reconstructed face, we partition a face into multiple subregions. 
This partitioning follows the face anatomy and each subregion contains one or several extended landmarks. 
A subregion will be revised if its associated extended landmarks are far away from the reconstructed face. 
To construct this partitioning, we follow the algorithm suggested in \cite{salazar20103d}, which defines a set of features on face using geometric curvatures and symmetry, then perform a stable tracing algorithm to obtain the segmentation. 
We perform and refine this segmentation on a template face, then transfer it to all other faces.
Fig.~\ref{Fig:Segmentation} illustrates the face segmentation on an example face model.

\subsection{Face Generator}
	
We build a face generator using a Generative Adversarial Network (GAN), and simultaneously train a generator $G$ and a discriminator $D$. 
$G$ maps a latent variable $z$, sampled from the prior distribution $p_z$, to a face image $G(z)$. 
$z$ is usually made a random vector (i.e., $p_z$ is uniform), and $z$ often has larger dimension than the dimension of image. 
In other words, the latent space is usually much bigger than the space of real faces.

\begin{figure}[h]
		\centering
		\begin{tabular}{cccc}
			\includegraphics[height=0.1\textheight] {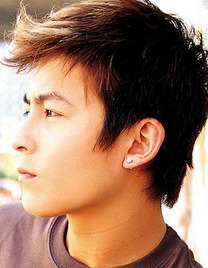} &
			\includegraphics[height=0.1\textheight] {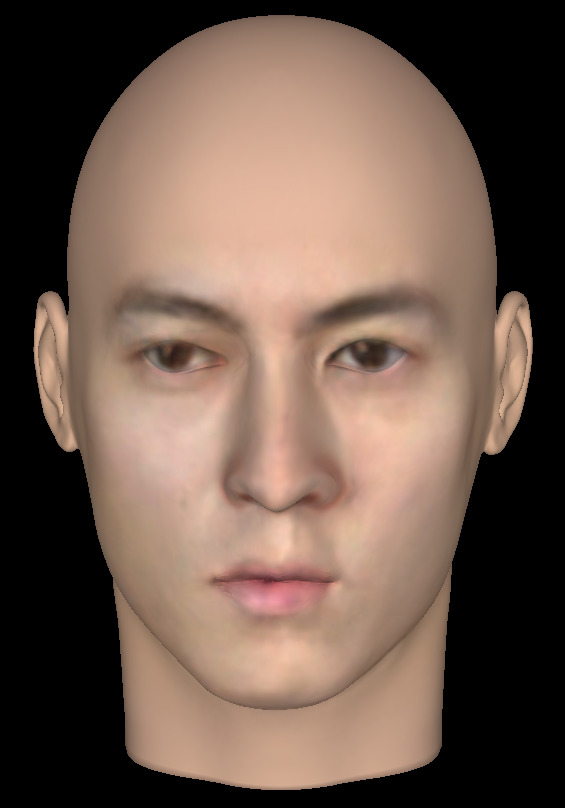} &
			\includegraphics[height=0.1\textheight] {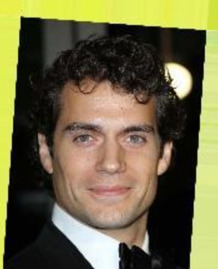} &
			\includegraphics[height=0.1\textheight] {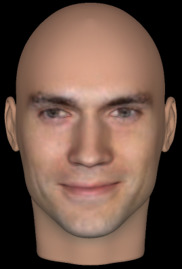} \\
			\includegraphics[height=0.1\textheight] {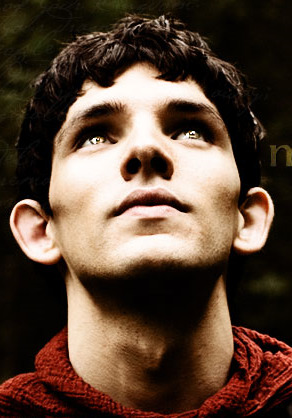} &
			\includegraphics[height=0.1\textheight] {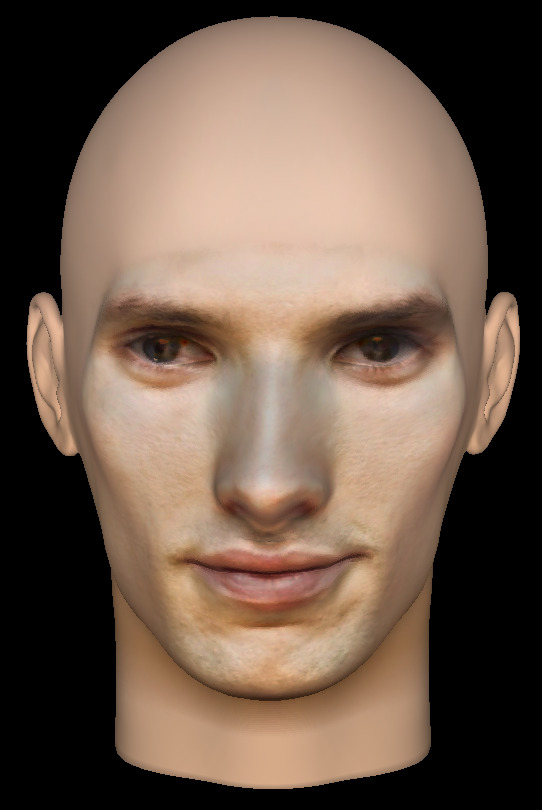} &
			\includegraphics[height=0.1\textheight] {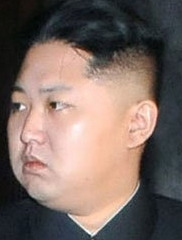} &
			\includegraphics[height=0.1\textheight] {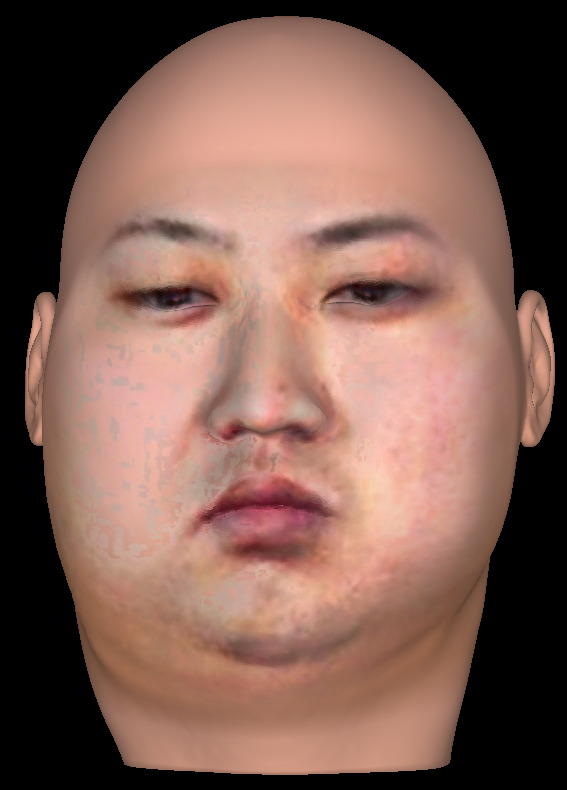} \\
		\end{tabular}
		\caption{Normalizing Face Images. Original photos in the odd columns are re-rendered and normalized to images in even columns.~\label{Fig:NormFace}}
	\end{figure}

\textbf{Face Image Normalization.} 
When training a GAN, it is known that using datasets with smaller distribution variance will usually lead to more efficient training and networks with better performance~\cite{roth2017stabilizing}. 
Furthermore, in this problem, our goal is to generate a good 3D face, rather than a realistic 2D face image.
Because of these two reasons, we train our GAN using rendered 3D faces, rather than using regular portrait photos. 
For each face image in our database, we first reconstruct its 3D face, then use a canonical camera pose and illumination configuration to render a new face image (see Section~\ref{Subsec:Decoder}). These normalized images will reduce the variance of distribution of the training images and can desirably improve the performance of the GAN. 
Some image normalization results are shown in Fig.~\ref{Fig:NormFace}.
	
By training a GAN with this normalized face dataset, we have a generator $G$ that can synthesize an image that looks like an image rendered from a real 3D face, and a discriminator $D$ that examines whether an image is from a real 3D face or a fake 3D face.

\subsection{Geometrically Constrained Generative Face Image Inpainting} 

When we use a GAN to generate a face image, after training,  feasible latent variables are from a high dimensional manifold $M$ in the latent space. 
Inpainting the missing part following its surrounding non-missing region reduces to restricting $G(z)$ in these non-missing regions.
In other words, this also restricts $z$ to be within some certain subspace $V$. If $M \bigcap V = \emptyset$, then we need to sacrifice some authenticity (face being less realistic) or context preservation (deviation on non-missing region) and find a $z$ to minimize a pre-defined loss error. 
If $M \bigcap V != \emptyset$, very often, there are many feasible solutions. 
In most existing GAN-based face generators, the latent space is usually very big and we observe that the latter case is what we often meet in practice. 
In other words, inpainting often has arbitrariness: many different generated faces satisfy the requirement and are equally good in certain sense. When the missing region is bigger, such arbitrariness or ambiguity becomes more significant. 

Therefore, here enforcing extra geometric constraints using extended landmarks defined on the skull not only is feasible, but also could effectively reduce the aforementioned ambiguity.

A state-of-the-art generative face model is introduced by Yeh et al.~\cite{yeh2017semantic}.
In this work, besides the commonly used \emph{context loss}, which  measures the deviation between the generated and given images on non-missing regions, a \emph{prior loss} is further considered to penalize the non-authenticity of the generated 2D face image. 
Because a trained GAN is not guaranteed to always produce fully realistic faces, having such a \emph{prior loss} could better control the authenticity of the generated face. In our work, as discussed above, we now incorporate an extra \emph{geometry loss} to reduce ambiguity and constrain the face inpainting following the given skull.

\textbf{Context Loss.}
To make the filled patch coherent with its surrounding contents, a \emph{context loss} is used to measure the deviation of $G(z)$ from the original image on these uncorrupted regions. A convenient metric is to use the $L_2$ norm.
But such a uniform measure equally considers all the pixels, which may not be desirable: we may want to pay most attention to pixels 
near the missing region, and not worry about the difference in the background. 
With this intuition, we define the importance of an uncorrupted pixel to be positively correlated with the number of its surrounding corrupted pixels. Then, a pixel that is far away from any hole has small importance and plays little role in the inpainting process. This importance weighting term $W_i$ at pixel $i$ is then defined as 
\[
W_i = \left\{ 
\begin{array}{ll}
\frac{1}{|N(i)|} \sum_{j \in N(i)} (1-B_j) & , \; \text{if} \; B_i =1 \\
0 & , \; \text{if} \;  B_i = 0
\end{array}
\right.
\]
where the mask pixel $B_i=0$ ($M_i =1$) means pixel $i$ is missing (not missing), $N(i)$ refers to the set of neighboring pixels of $i$ in a local window, and $|N(i)|$ denotes the cardinality of $N(i)$.
Empirically, we found the $L_1$-norm to perform
slightly better than the $L_2$-norm in our framework. Taking it
all together, we define the context loss to be a weighted
$L_1$-norm difference between the recovered image and the uncorrupted portion, 
\begin{equation}
\mathcal{L}_c(z|y, B) = ||W \circ (G(z) - y)||_1,
\end{equation}
where $\circ$ denotes the element-wise multiplication.

\textbf{Prior Loss.}
The prior loss was introduced as a class of penalties based on high-level image feature representations instead of pixelwise differences. 
It encourages the recovered image to be similar with the samples drawn from the training set, and penalizes an unrealistic generated image. 
The discriminator $D$ in the GAN is trained to differentiate generated images from real images. 
Therefore, we can directly use $D$ to help define the prior loss, i.e., 
\begin{equation}
\mathcal{L}_p(z) = \lambda_p \log(1 - D(G(z))),
\end{equation}
where $\lambda_p$ is a weight parameter for prior loss. 
The intuition of introducing the prior loss is that, when mapping a corrupted image to a point $z$ in the latent space, we require this $z$ to not only have small context error, but also can produce an authentic face, or in other words, $G(z)$ should be able to fool $D$. 
We use the value of prior loss to numerically measure the authenticity of the face generated by our GAN.

\textbf{Geometry Loss.}
To further reduce the face ambiguity following the skull, we extend the anthropometric landmarks defined on the given skull following statistically measured tissue depths on these regions, then use these extended landmarks to constrain the geometry of synthesized face.  
Specifically, let $S$ be the set of extended landmark points. The  corresponding feature vertices on the generated face should pass the extended landmarks. 
Furthermore, these anthropometric landmarks are mainly defined on the frontal face below the eyes. 
To remove arbitrariness on other regions such as forehead and the back side of the head, we also define depth guidance on these regions. We call these regions the \emph{definite region}. 
According to \cite{phillips1996facial}, the tissue depths are often considered constant on these regions.
	
The landmark index-$1$ in Fig.~\ref{Fig:Landmark} (a) is in the forehead definite region, we use the tissue depth $d_1$ defined on this landmark to define the depth of this definite region. 
We create an offset $(\delta=d_1)$ surface $\tilde{S}$ from the querying skull, then project the vertices in the definite region of the face onto $\tilde{S}$. These projected vertices will be treated as new extended landmarks. Other definite regions such as the back side of the head are processed similarly. 

Given a parameter $z$ and its reconstructed face image $G(z)$, let $\Psi_E$ be the face encoder and $\Psi_E(G(z))=(\alpha, \delta, \theta)$ be the semantic code, which can be used to reconstruct the 3D face using the FLAME model, $M(\Psi_E(G(z)))$. 
The Geometry Loss is defined as
	\begin{equation}
	\mathcal{L}_g(z|S) = \lambda_2 \sum_i (\|L_f(M(\Psi_E(G(z))))_i 
	- L_s(S)_i\|)
	\end{equation}
	where $\lambda_2$ is a weight parameter for geometry loss, $S$ is the skull, and $L_f()_i, L_s()_i$ are the $i$-th feature points on the face and $i$-th extended landmarks on the skull, respectively.

\textbf{Composed Objective Function.}
Combining the above three terms, we define the final objective function. 
Given a corrupted image $y$ and a binary mask $B$ indicating the missing region on $y$, the inpainting reduces to solving 
\begin{equation}
\hat{z} = \arg \min_{z}{\mathcal{L}_c(z|y,B) + \mathcal{L}_p(z) + \mathcal{L}_g(z|S)},
\label{Eqn:hat_z}
\end{equation} 
where $\mathcal{L}_c$ is the \emph{context loss} measuring deviation on non-corrupted region, $\mathcal{L}_p$ is the \emph{prior loss} evaluating the authenticity of $G(\hat{z})$, and	$\mathcal{L}_g$ is the \emph{geometry loss} penalizing superimposition errors. 
	
\begin{figure}[h]
		\centering
		\begin{tabular}{cccc}
			\includegraphics[height=0.11\textheight] {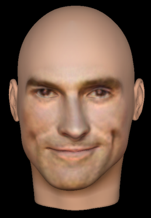} &
			\includegraphics[height=0.11\textheight] {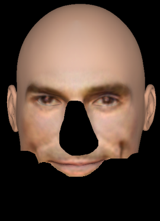} &
			\includegraphics[height=0.11\textheight] {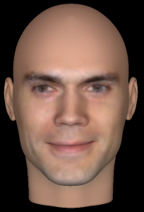} &
			\includegraphics[height=0.11\textheight] {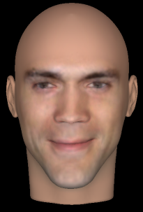} \\
			(a) & (b) & (c) 0.84/-4.41/0.74 & (d) 0.87/-4.45/0.61 
		\end{tabular}
		\caption{Inpainting with only context loss and prior loss. The original reconstructed face (a) has its non-matched region removed (b) for re-synthesis. If only context and prior losses are used, the two inpainted faces (c, d) have similar context and prior loss, but different geometry loss. The three numbers in sub-captions are context, prior, and geometry loss, respectively.~\label{Fig:InpaintingAmbiguity}}
	\end{figure}

Introducing the geometry loss into our model allows the skull to guide the face inpainting and it effectively reduces the arbitrariness of GAN generated faces. An example is illustrated in Fig.~\ref{Fig:InpaintingAmbiguity}.  
In Fig.\ref{Fig:InpaintingAmbiguity} (c, d), the two faces have similar prior and context losses. Without geometry loss, they are equally good according to the generator. This ambiguity can be reduced with the help of the skull after integrating geometry loss. 

	\begin{figure*}[h]
		\centering
		\begin{tabular}{c c c c p{-0.01cm} c p{-0.01cm} c p{-0.01cm} c c}
			\includegraphics[height=0.11\textheight] {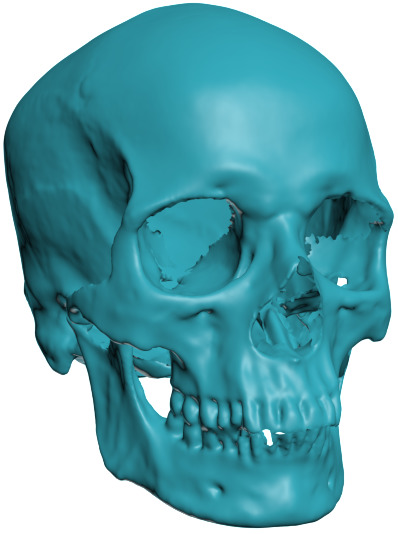} &
			\includegraphics[height=0.11\textheight] {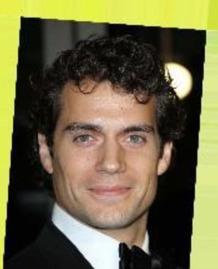} &	
			\includegraphics[height=0.11\textheight] {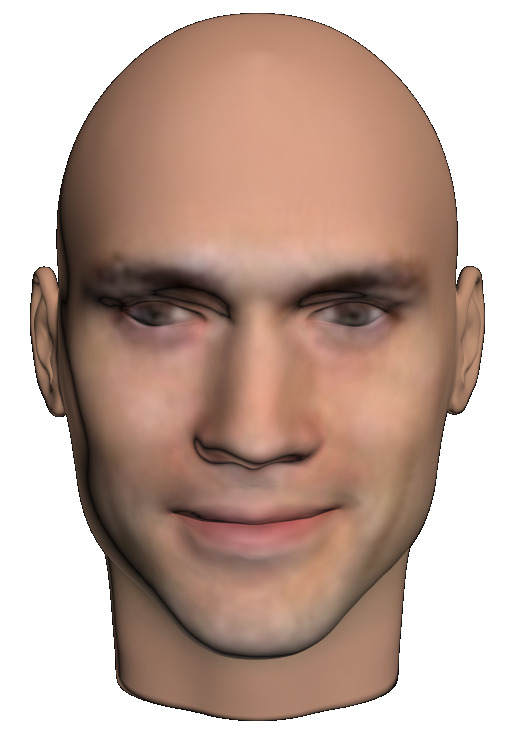} &
			\includegraphics[height=0.11\textheight] {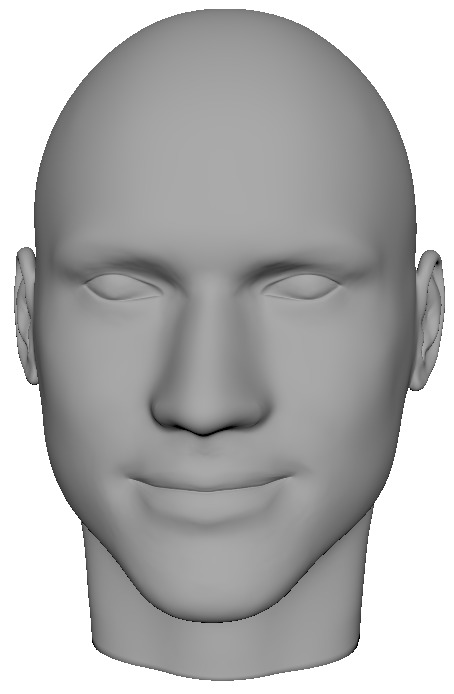} & &
			\includegraphics[height=0.11\textheight] {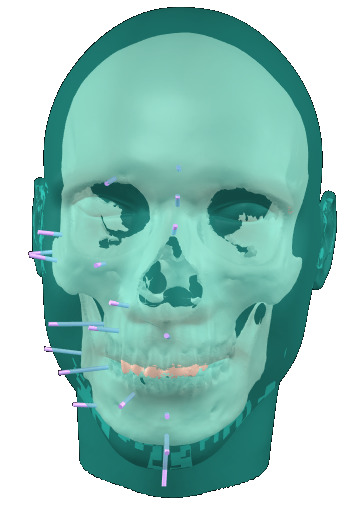} & &
			\includegraphics[height=0.11\textheight] {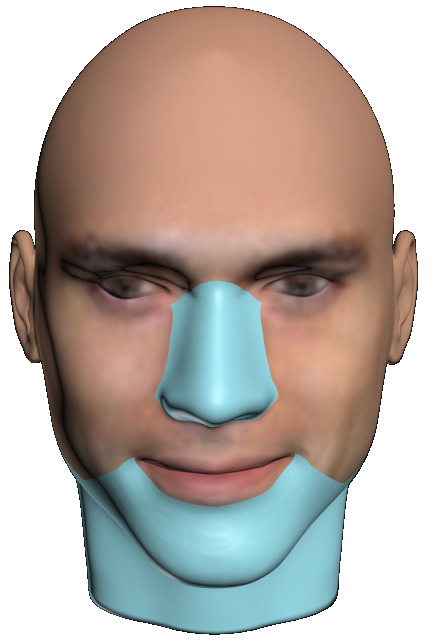} & &
			\includegraphics[height=0.11\textheight] {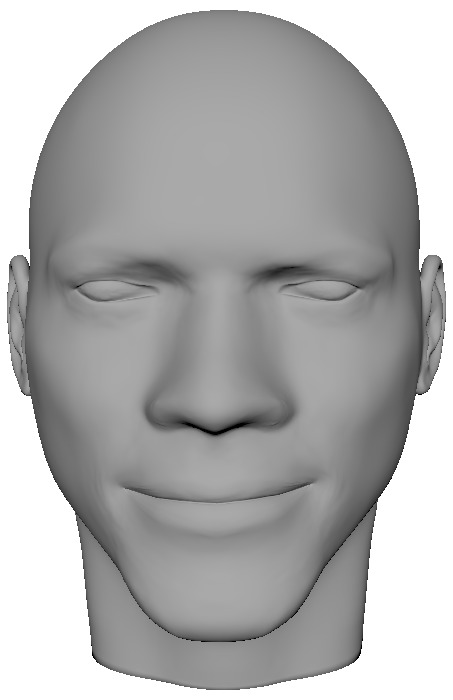} &
			\includegraphics[width=0.11\textheight] {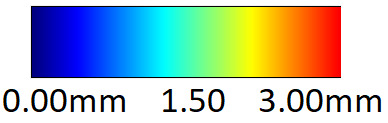} \\
			(a) & (b) & (c) & (d) && (e) && (f) && (g) 0.132/-4.43/0 &  \\			
			\includegraphics[height=0.11\textheight] {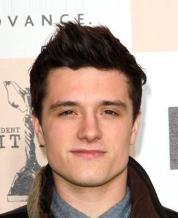} &
			\includegraphics[height=0.11\textheight] {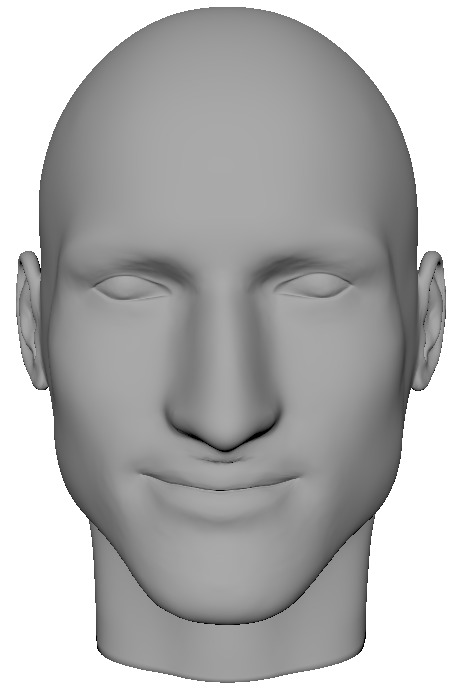} &
			\includegraphics[height=0.11\textheight] {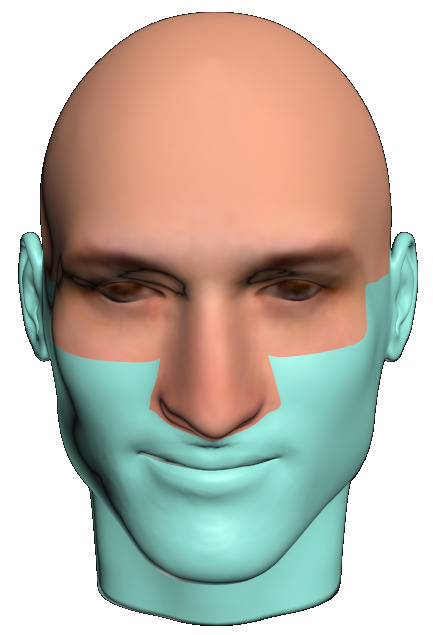} &
			\includegraphics[height=0.11\textheight] {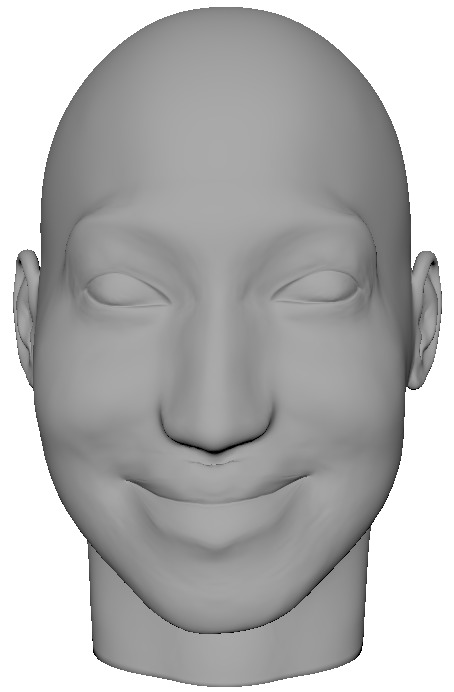}& \includegraphics[height=0.11\textheight] {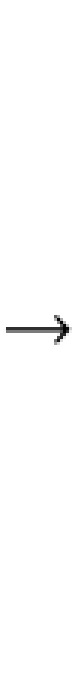} &
			\includegraphics[height=0.11\textheight] {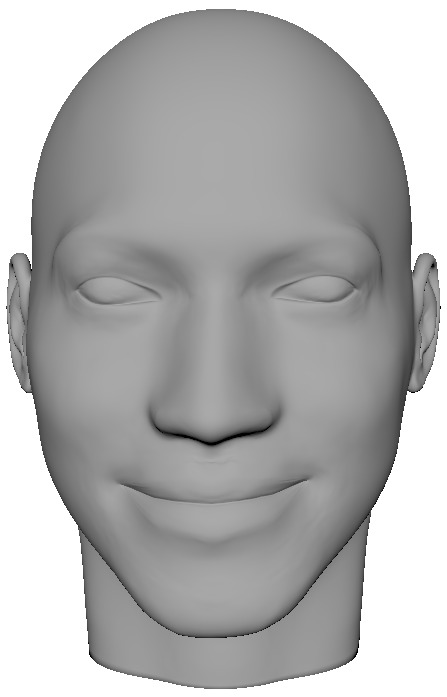}& \includegraphics[height=0.11\textheight] {figs/inpainting-progress/arrow} &
			\includegraphics[height=0.11\textheight] {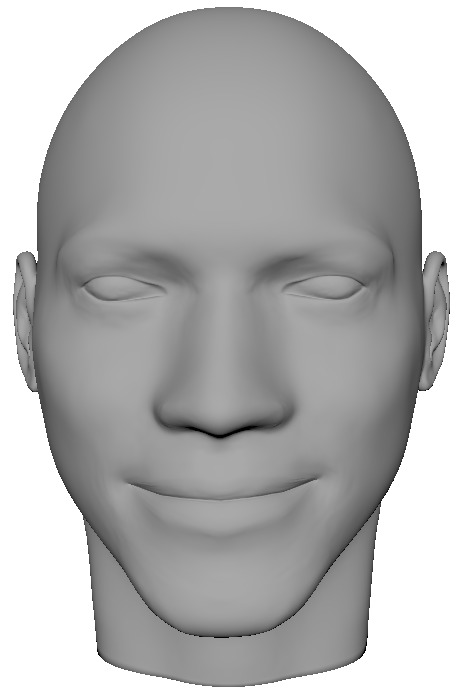}& \includegraphics[height=0.11\textheight] {figs/inpainting-progress/arrow} &
			\includegraphics[height=0.11\textheight] {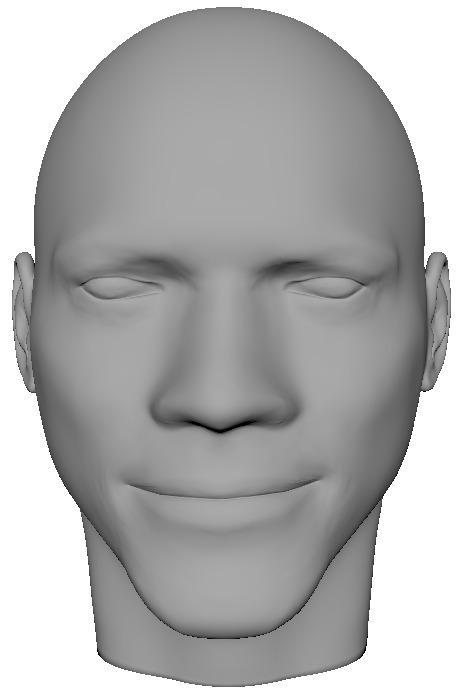}&
			\includegraphics[height=0.11\textheight] {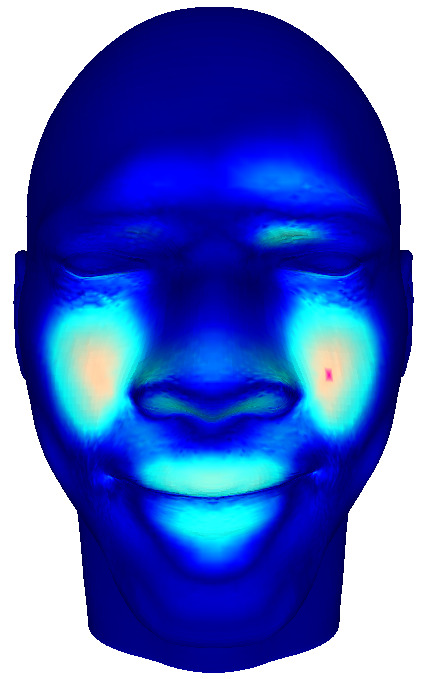}\\
			(h) & (i) & (j) & (k) && (l) && (m) && (n) 0.133/-4.29/0 & (o) \\
			\includegraphics[height=0.11\textheight] {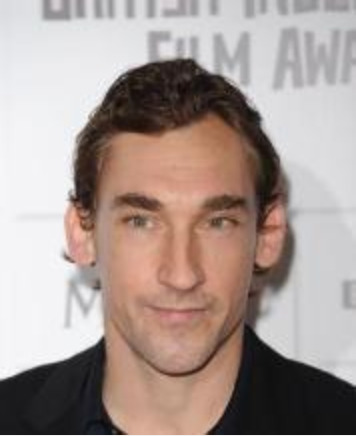} &
			\includegraphics[height=0.11\textheight] {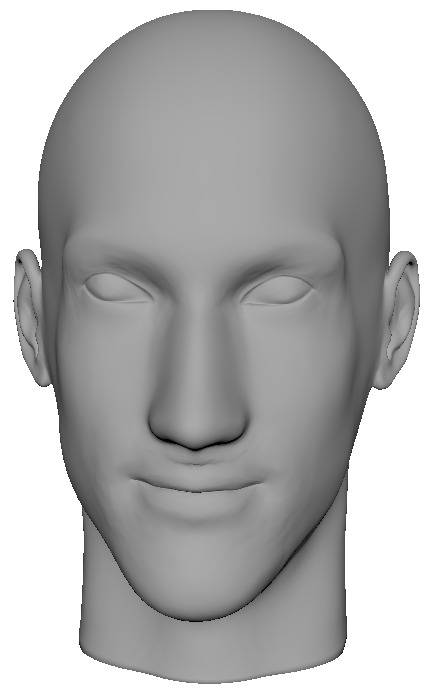} &
			\includegraphics[height=0.11\textheight] {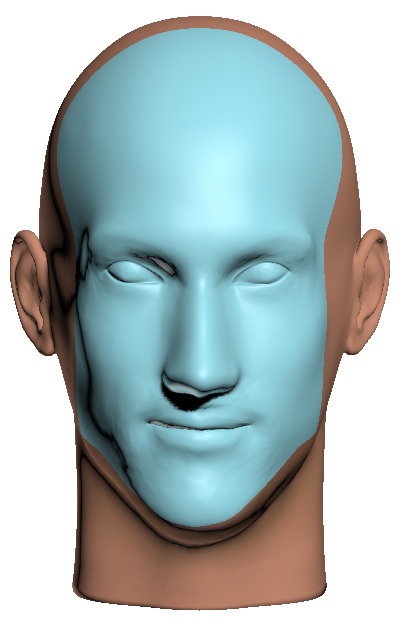} &
			\includegraphics[height=0.11\textheight] {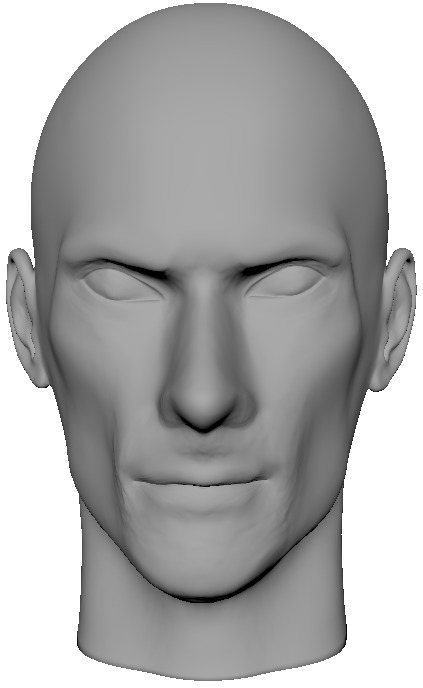} & \includegraphics[height=0.11\textheight] {figs/inpainting-progress/arrow} &
			\includegraphics[height=0.11\textheight] {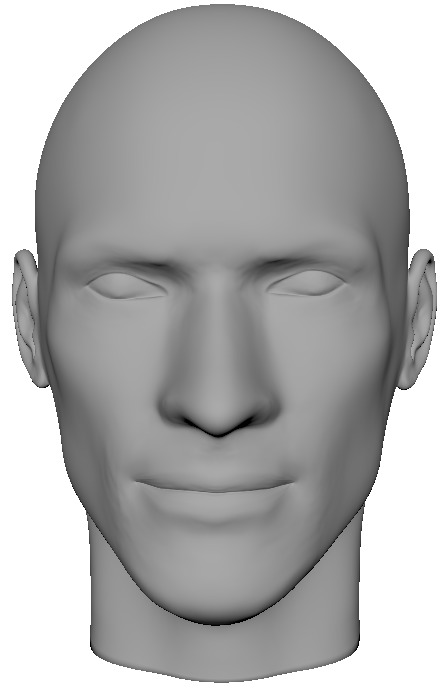}& \includegraphics[height=0.11\textheight] {figs/inpainting-progress/arrow} &
			\includegraphics[height=0.11\textheight] {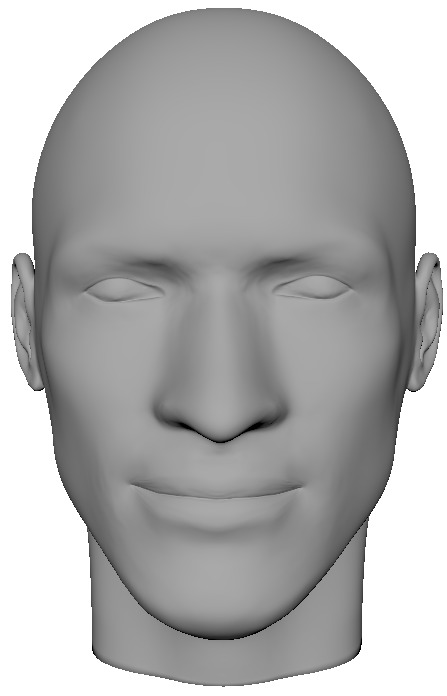}& \includegraphics[height=0.11\textheight] {figs/inpainting-progress/arrow} &
			\includegraphics[height=0.11\textheight] {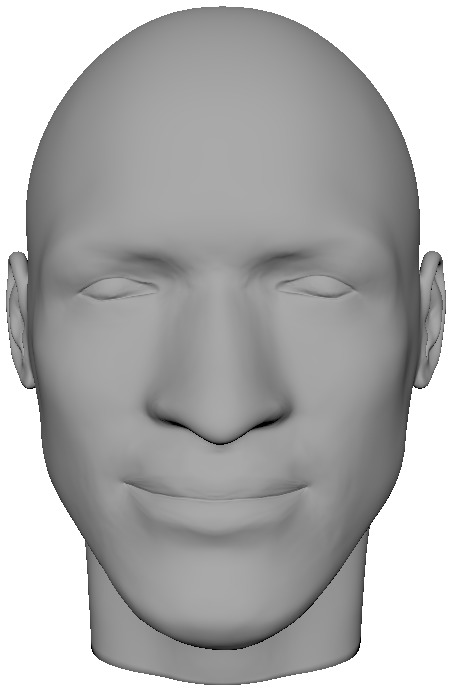}&
			\includegraphics[height=0.11\textheight] {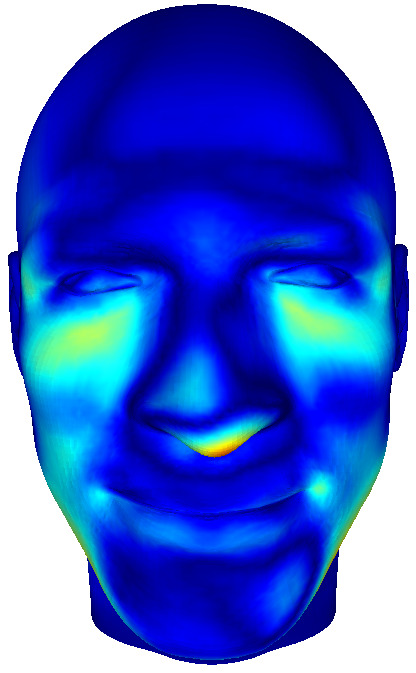}\\
			(p) & (q) & (r) & (s) && (t) && (u) && (v) 0.178/-4.31/0 & (w) \\
		\end{tabular}
		
		\begin{tabular}{c}
			\includegraphics[height=0.26\textheight] {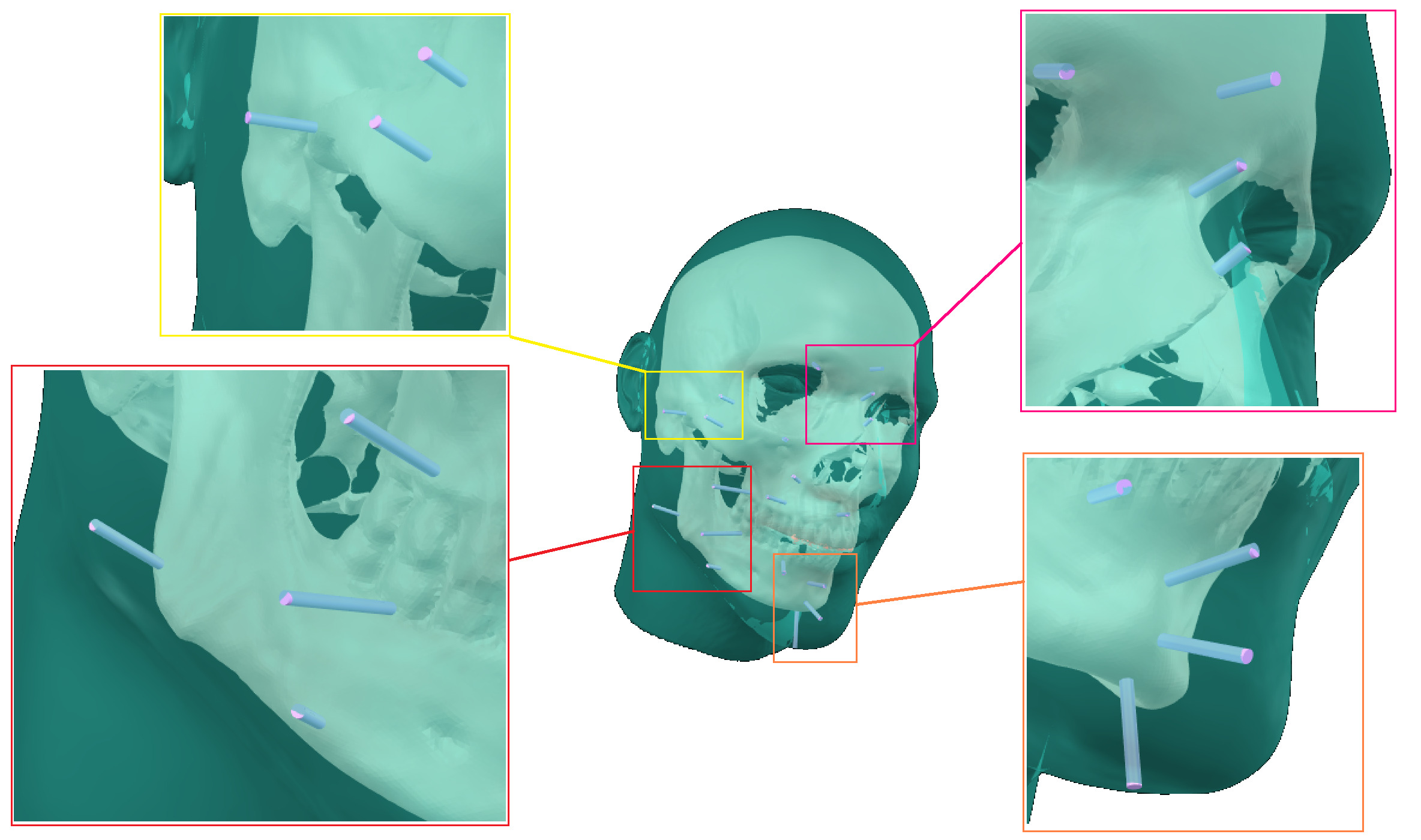}  \\
			(x) \\
		\end{tabular}
		\caption{Stability and Accuracy of Our Inpainting Result. To reconstruct a face that matches with the querying skull (a), we start from three different faces (b, h, p), whose reconstructed 3D faces are (c,i, q). The initial superimpositions suggest different modification regions (f, j, r): cyan regions are not well-matched and will be removed and re-synthesized. After inpainting we obtain similar resultant faces (g, n and v). The maximal and average point-to-point deviations between (g) and (n) are $2.1mm$ and $0.5mm$ respectively, as color-encoded in (o). And the maximal and average point-to-point deviations between (g) and (v) are $2.9mm$ and $1.3mm$ respectively, as color-encoded in (w). The sub-captions of (g, n, v) give the context loss/prior loss/geometry loss values. The reconstructed face passes all the extended landmarks, as shown in the zoom-in view (x). 
		In the second row, in (k), starting from a random initial $z$, $G(z)$ gives the initial face. (l, m) are the intermediate stages of the optimization before it converges to (n). 
		Similarly, in the third row, (s, t, u, v) shows the refinement of face (q) during our optimization.~\label{Fig:SideBySideComp}}
	\end{figure*}

\begin{figure}[h]
		\centering
		\begin{tabular}{cc}
			\includegraphics[height=0.17\textwidth] {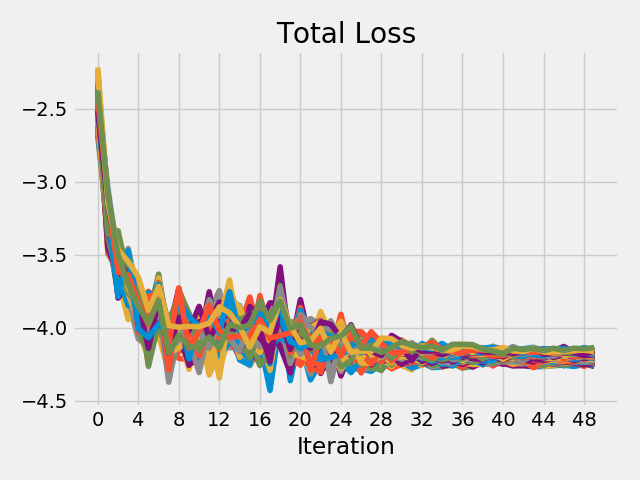} &
			\includegraphics[height=0.17\textwidth] {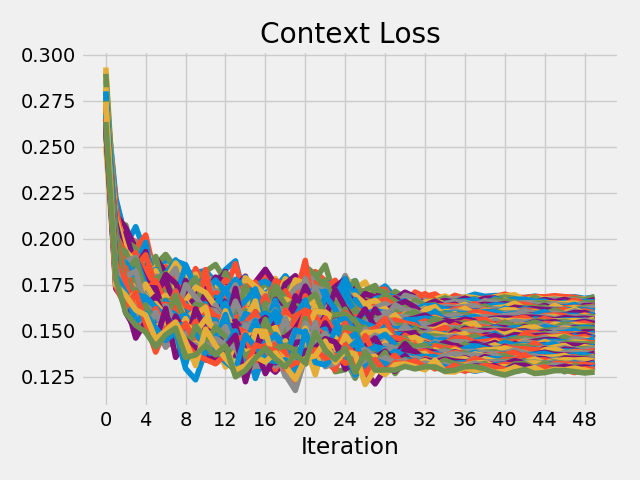} \\
			(a) & (b) \\
			\includegraphics[height=0.17\textwidth] {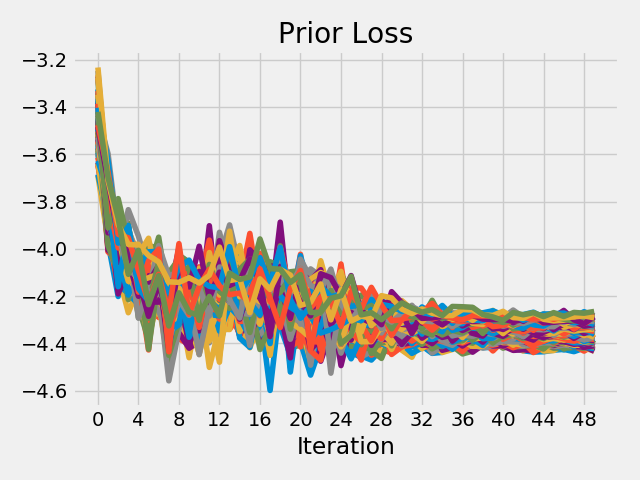} &
			\includegraphics[height=0.17\textwidth] {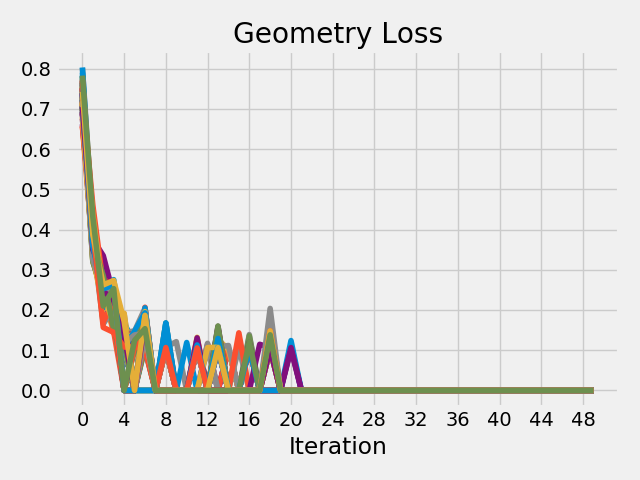} \\
			(c) & (d) \\
		\end{tabular}
		\caption{
			The evolution of the losses during optimization, we plot the curves with two different starting face images.(a) the total loss curve; (b) the context loss curve, the dashed curve are the context loss without geometry loss; (c) the prior loss curve; (d) the geometry loss curve.~\label{Fig:LossCurve}}
	\end{figure}

	Given a corrupted image, we can find an optimal latent variable $\hat{z}$ in the latent representation space by minimizing the total loss composed of the aforementioned context, prior and geometry loss terms. We use a randomly generated $z$ as the initial variable and iteratively refine it through back-propagation until it converges.  
	In Fig.~\ref{Fig:LossCurve}, we plot the convergence of the composed objective function and the three separate loss terms. The experiments were conducted on $100$ different initial candidate faces for the query skull in Fig.~\ref{Fig:SideBySideComp}(a). We can see that optimization converges within about $50$ iterations. 
	
\subsection{Analysis on Accuracy and Stability of Our Inpainting}
\label{Subsec:Stability}

With the trained GAN, the inpainting reduces to searching for the optimal latent variable in the latent space. 
With $\mathcal{L}_c$, we reduce the search to a subspace $p^{c}_{z}$ from which latent variables will generate images that have similar content with $y$ in the non-missing regions. 
But $G(z^{c})$ is not guaranteed to be always realistic, and introducing $\mathcal{L}_p$ helps further reduce the search space $p^{c}_z$ to $p^{c,p}_z$, in which latent variables are producing realistic faces.

With $\mathcal{L}_c$ and $\mathcal{L}_p$, the generated image will look like a real face, and such a result is what \cite{yeh2017semantic} suggests. However, when the removed region on the face is big, the inpainted result does have some ambiguity and randomness. Because $\mathcal{L}_c$ does not apply on these missing regions, whose repair is determined by $L_p$ (i.e. the discriminator $D$). But $D$ tolerates all realistic images and reports a small error when $G(z)$ looks realistic. With a good enough $G$, the acceptable range of $z$ is still quite big. We found that its variance is still big, and the $G(z)$ could be quite random: the faces can be similar on the uncorrupted regions, but have different appearances on the corrupted regions.  

Therefore, we enforce the face to have a good superimposition with the querying skull by introducing $\mathcal{L}_g$, and further reduce the search space from $p^{c,p}_z$ to $p^{c,p,g}$. 
$\mathcal{L}_g$ has strong control on the face appearance on the missing region, and introduces very big cost when the face geometry deviates from the constraints of the skull.
With $\mathcal{L}_g$, the randomness of face appearance greatly reduces, and $G(z)$ becomes quite stable and controllable. 

Fig.~\ref{Fig:SideBySideComp} illustrates an inpainting example that shows the stability of our algorithm.
Given a querying skull (a), although we start from three different candidate faces (see their original photos (b,h,p), reconstructed faces (d,i,q), respectively). In (e) we show the distribution of the superimposition error of the first candidate, which has quite large error on some regions. After the face inpainting, the reconstructed faces (g), (n) and (v) are very similar, as shown in the color-encoded face (o) and (w). This shows the stability of our algorithm.

Fig.~\ref{Fig:SideBySideComp}(o) encoded the vertex-to-vertex deviation between faces in (g) and (n). We can see most of the regions have similar geometry, except some regions around the cheek have larger deviation, this is caused by the arbitrariness we discussed in Sec.~\ref{Sec:FaceResynthesis}. The maximal (average) vertex-to-vertex deviation between faces on frontal area in (g) and (n) is $2.1mm$ ($0.5mm$), and for (w) which encoding the vertex-to-vertex deviation between faces in (g) and (v), the deviation is $2.9mm$ ($1.3mm$). Note that those values are significantly smaller than the face reconstruction error of $9.7mm$ ($2.2mm$) in Table.~\ref{Tab:ReconComp}. So we can consider these two reconstructed faces to be nearly identical, especially in frontal area. This demonstrates that our approach can generate stable result. 

From the context loss/prior loss/geometry loss values under Fig.~\ref{Fig:SideBySideComp}(g), (n) and (v), we see that they all have low prior loss and $0$ geometry loss. 
The superimposition error is totally eliminated, indicating that the reconstruction is realistic and matches the skull very well. An observation is they have difference context losses. This is because the optimization is globally updated, optimizing the geometry loss will change the non-missing region. Sometimes, there is a trade off between the. Here since we emphasize in reducing (using bigger weight for) the geometry loss, some context loss needs to compromise. On the other hand, the context loss is evaluated based on the difference between the generated face and candidate image. Since the candidate images are different, our generated face will remain some features from the image, especially on the sparse landmark regions. Fig.\ref{Fig:SideBySideComp}(k-n) and (s-v) show how a randomly starting face gradually changing in the optimization progress.

	\section{Experimental Results}
	\label{Sec:Results}

	\subsection{Image-based Face Reconstruction}
	\begin{figure*}[h]
		\centering
		\begin{tabular}{cccccc}
			& P3DMM & MOFA & F3DMM & PFMOFA & GFMOFA (Ours)\\
			\includegraphics[width=0.15\textheight, valign=m] {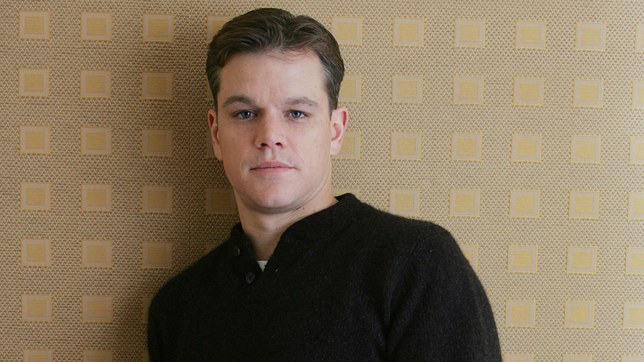} &
			\includegraphics[height=0.12\textheight, valign=m] {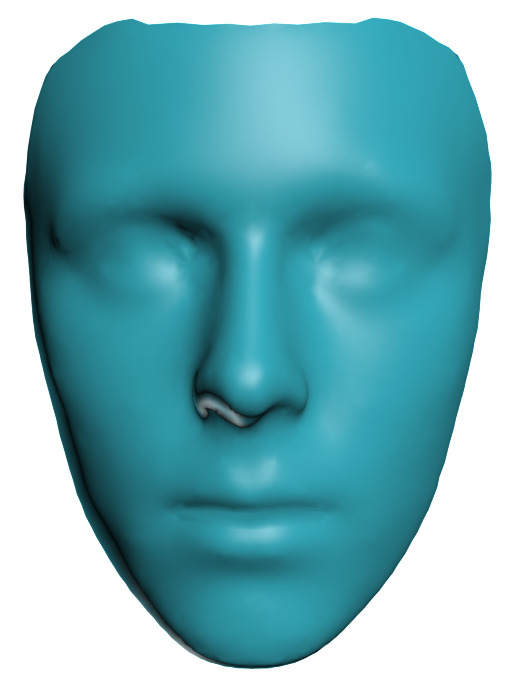} &
			\includegraphics[height=0.12\textheight, valign=m] {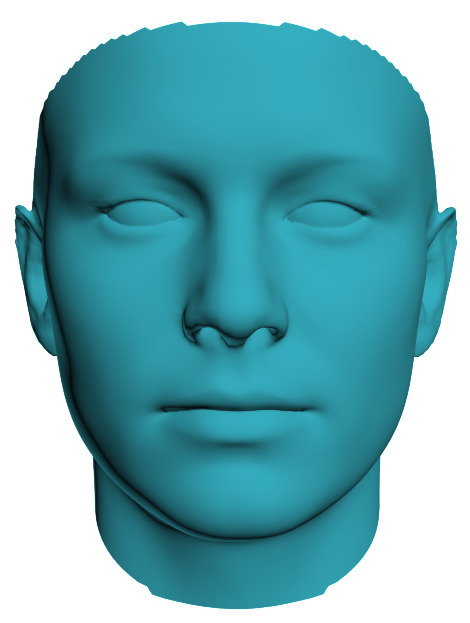} &
			\includegraphics[height=0.12\textheight, valign=m] {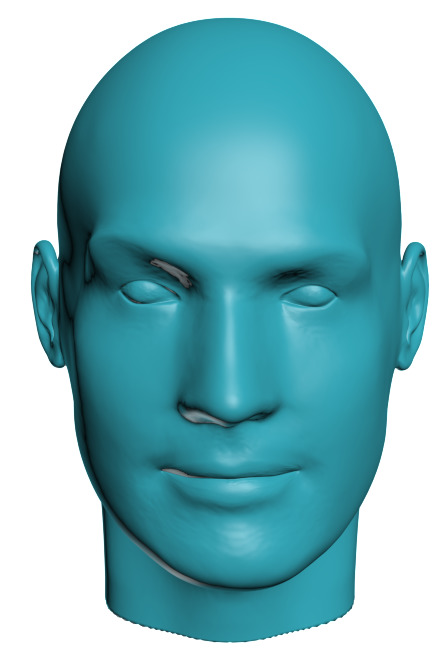} &
			\includegraphics[height=0.12\textheight, valign=m] {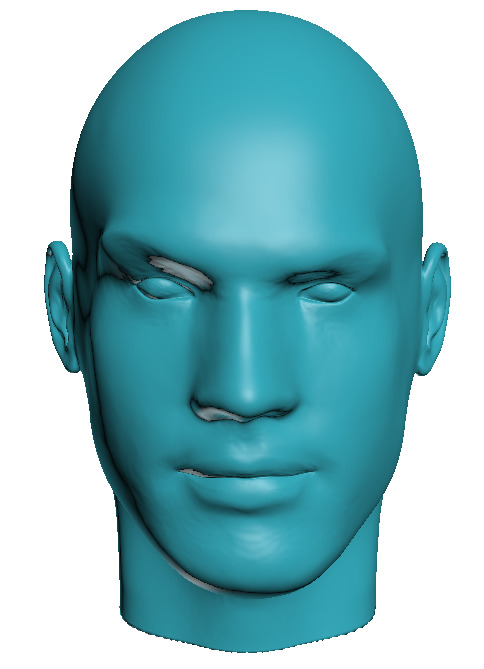} &
			\includegraphics[height=0.12\textheight, valign=m] {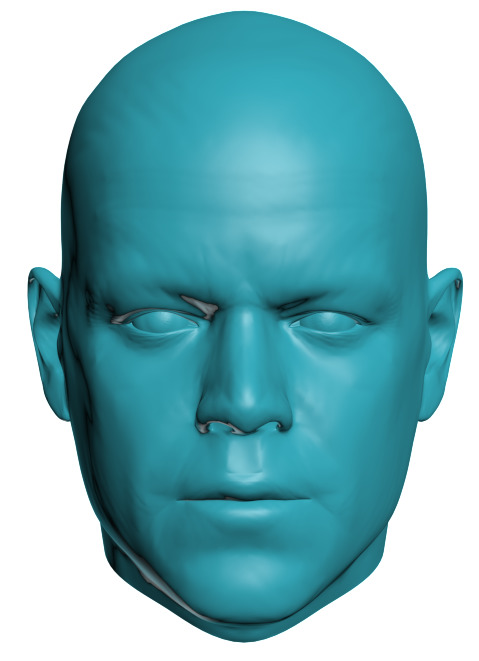} \\
			
			\includegraphics[width=0.15\textheight, valign=m] {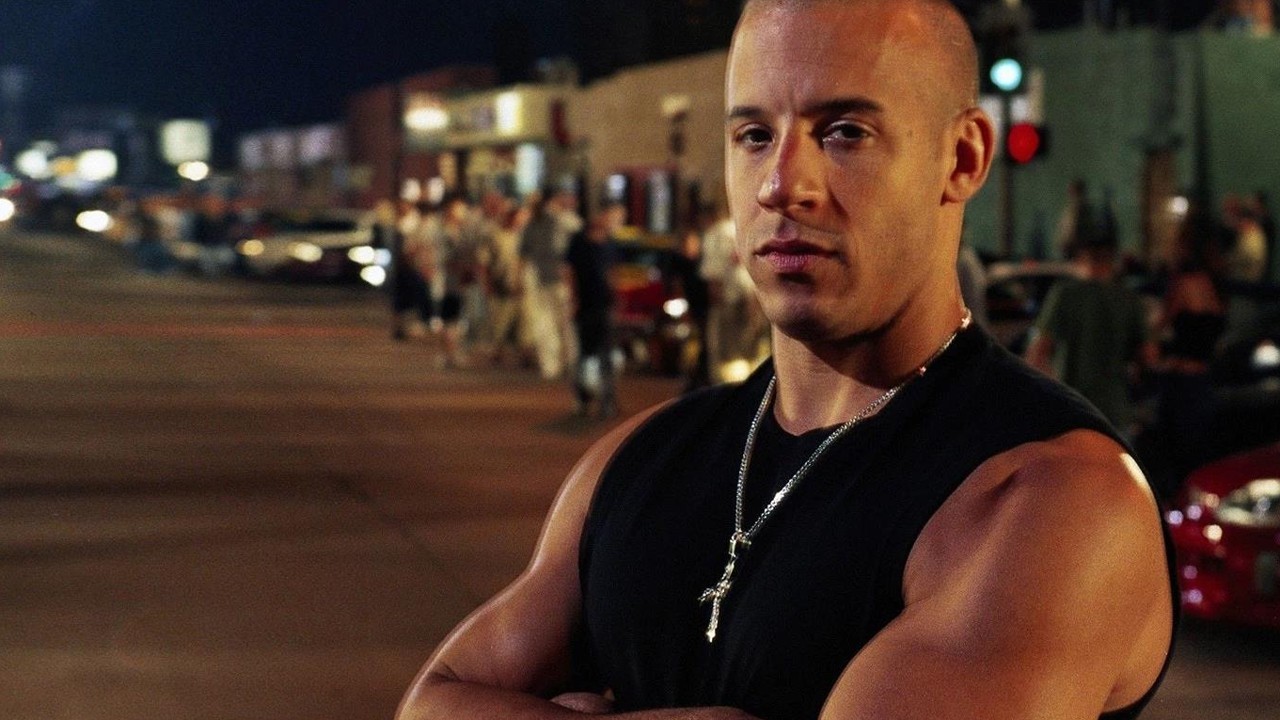} &
			\includegraphics[height=0.12\textheight, valign=m] {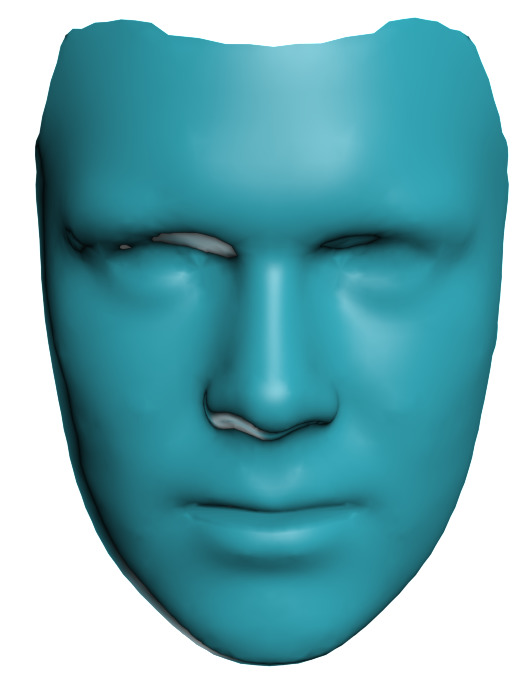} &
			\includegraphics[height=0.12\textheight, valign=m] {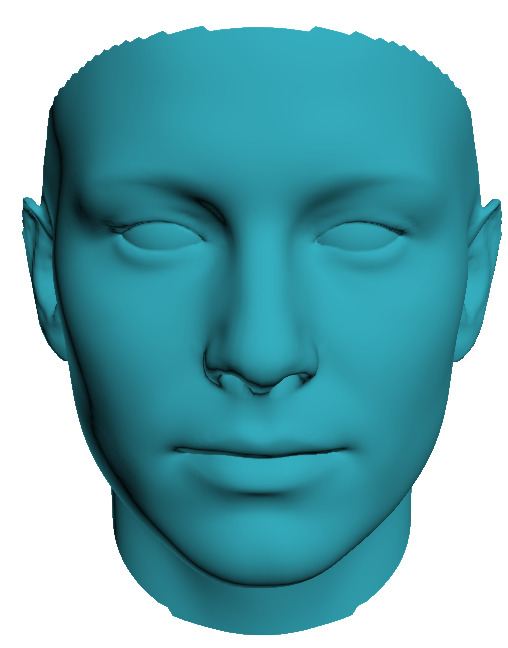} &
			\includegraphics[height=0.12\textheight, valign=m] {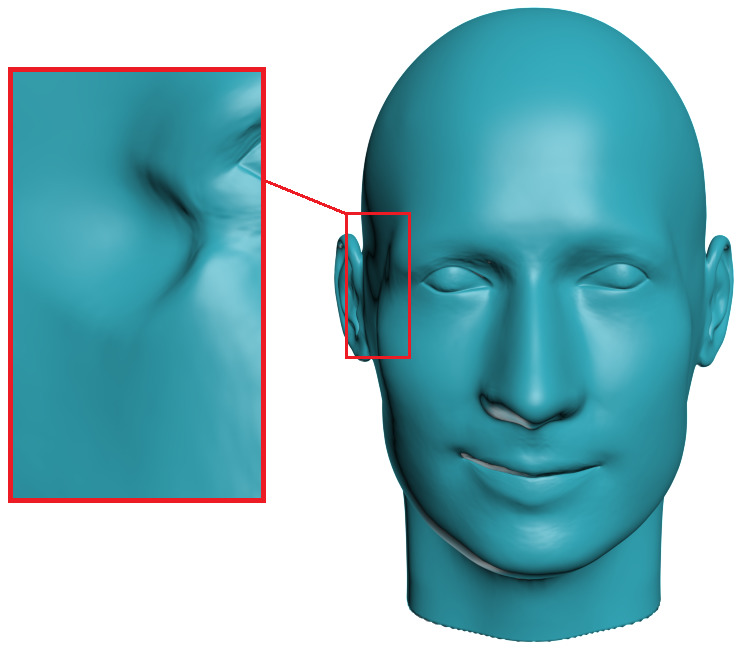} &
			\includegraphics[height=0.12\textheight, valign=m] {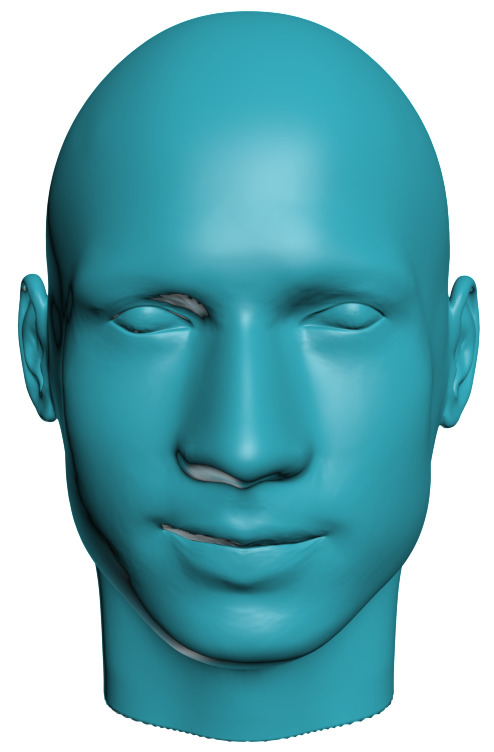} &
			\includegraphics[height=0.12\textheight, valign=m] {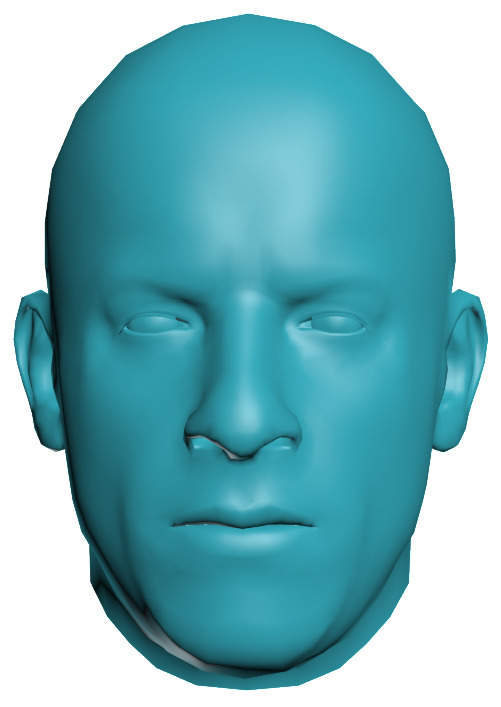} \\
			
			\includegraphics[width=0.15\textheight, valign=m] {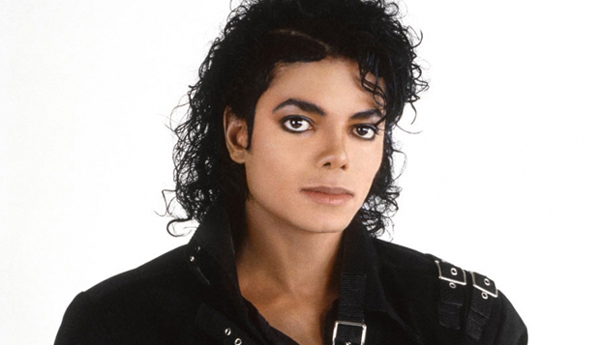} &
			\includegraphics[height=0.12\textheight, valign=m] {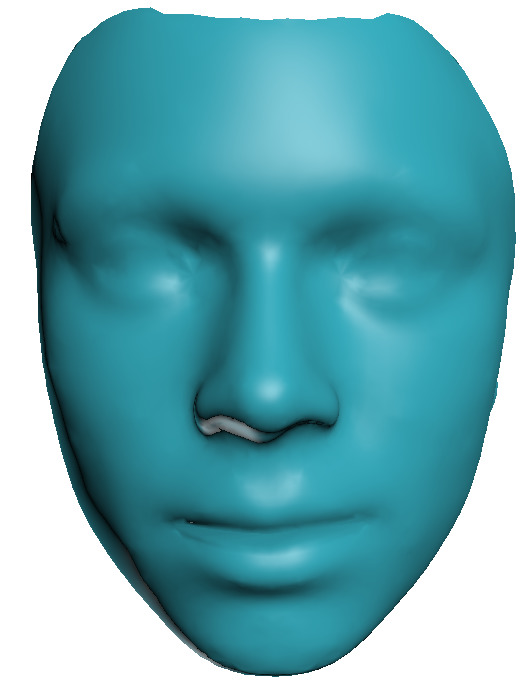} &
			\includegraphics[height=0.12\textheight, valign=m] {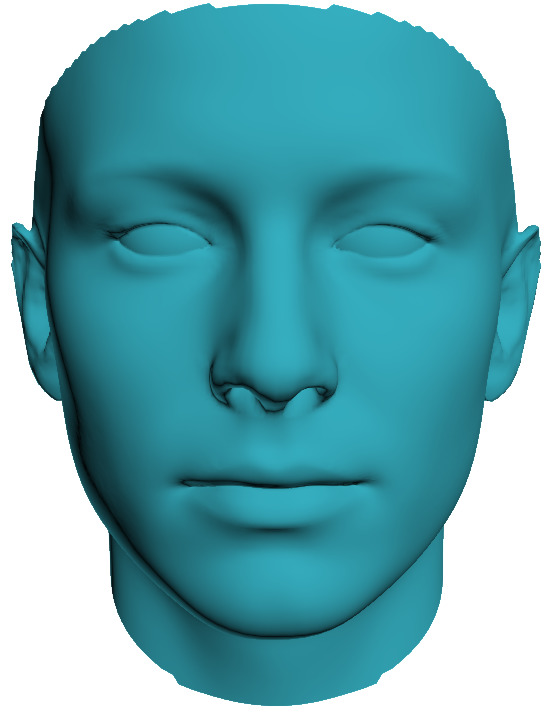} &
			\includegraphics[height=0.12\textheight, valign=m] {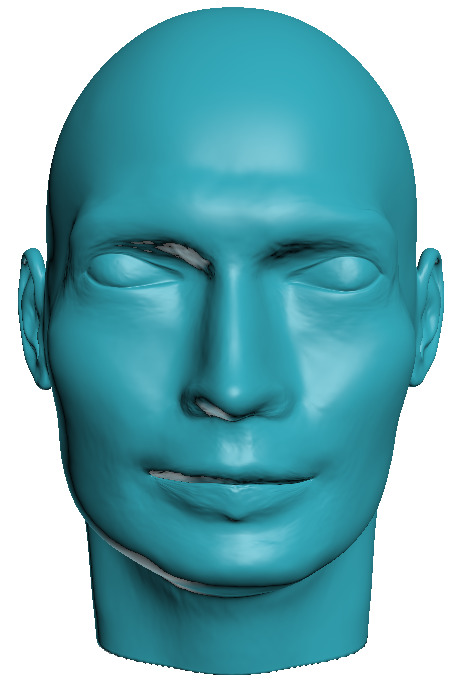} &
			\includegraphics[height=0.12\textheight, valign=m] {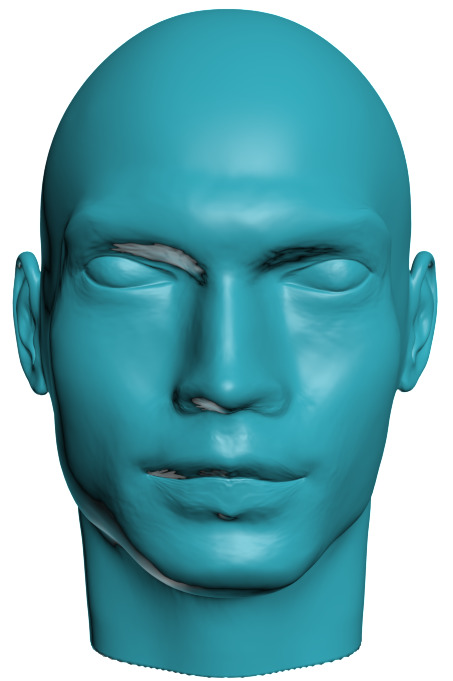} &
			\includegraphics[height=0.12\textheight, valign=m] {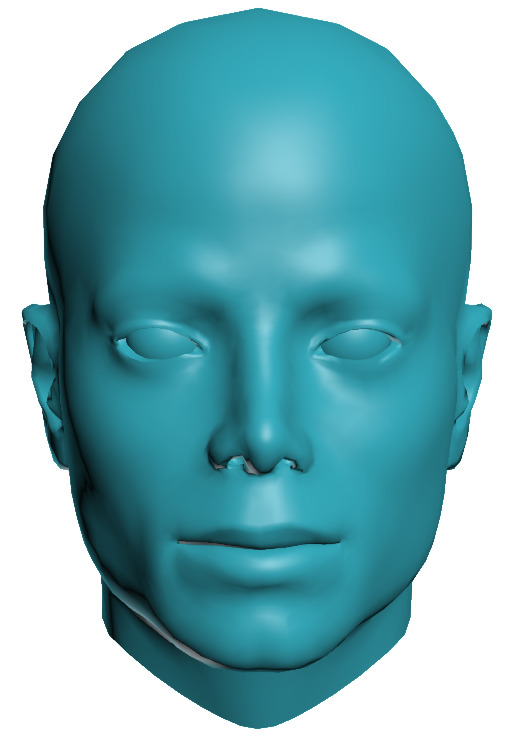} \\
		\end{tabular}
		\caption{Comparisons of Image-based Facial Reconstruction results.~\label{Fig:ReconComp}}
	\end{figure*}
	
To demonstrate the effectiveness of our image-based face reconstruction, we compare our algorithm with other recent algorithms, including (1) PCA-based 3DMM~\cite{blanz1999morphable}, or P3DMM (the widely adopted template-based deformation strategy), (2) Model-based Face Autoencoder~\cite{tewari2017mofa}, or MOFA (the start-of-the-art PCA-based deep face autoencoder), (3) FLAME-based 3DMM, or F3DMM (where we fit a FLAME\cite{FLAME:SiggraphAsia2017} face model), and (4) Pixel-based FLAME-MOFA, or PFMOFA (where the FLAME model replaces the PCA in building the MOFA autoencoder). 
Our algorithm incorporates the FLAME model in the MOFA structure. But instead of using pixel-wise loss function, we use geometric distance variation of features as the loss
(See the our geometric loss defined in Eqn.~(\ref{Eqn:E_Loss}) and Eqn.~(\ref{Eqn:E_Loss_Mult})). Therefore, we also denote our algorithm as Geometry-based FLAME-MOFA, or \emph{GFMOFA}.
	
Fig.~\ref{Fig:ReconComp} illustrates facial reconstructions performed on several Internet face images. They allow us to \textbf{qualitatively} compare these methods.  We can see that P3DMM generates overly smoothed faces with certain details missing. 
MOFA generates faces stably. But by using PCA, it also results in somewhat similar faces with fine details lost.
The FLAME-based 3DMM enhances detail on reconstructed faces, but it may produce artifacts in occluded face regions (see the zoom-in figure in Row-2). 
When using Pixel-based FLAME-MOFA, the reconstructions are more stable even with occlusions. However, with the pixel-based loss, a majority portion of the parameters in the encoder's network is used to infer the rendering coefficients (which may affect pixel values more significantly than face geometry). As a result, the quality of face geometry is not as good as our result. 
Among all these compared algorithms, our Geometry-based FLAME-MOFA stably reconstructs faces that have the highest geometric accuracy and most fine details.

	\begin{table}[h]
		\centering
		\caption{Comparisons of average and maximal point-to-point reconstruction errors, tested on randomly selected $2000$ pairs of 3D face and its portrait image from database~\cite{Guo20183DFace}. \label{Tab:ReconComp}}
		\begin{tabular}{|c|c|c|}
			\hline
			& Mean error & Largest error \\ \hline
			3DMM             & 5.4mm      & 24.5mm        \\ \hline
			MOFA             & 4.2mm      & 15.8mm        \\ \hline
			F3DMM       & 3.8mm      & 12.5mm        \\ \hline
			PFMOFA & 2.7mm      & 10.5mm        \\ \hline
			Ours             & 2.2mm      & 9.7mm         \\ \hline      
		\end{tabular}
	\end{table}

We also perform \textbf{quantitative} evaluations on reconstructed faces. We used $2000$ random pairs of 3D face and its portrait image from database~\cite{Guo20183DFace} as the testing dataset. We computed the vertex-to-vertex deviation between each groundtruth face and the reconstructed face, and documented the average and maximal errors in Table~\ref{Tab:ReconComp}.
From these statistics, we can see that our GFMOFA algorithm produces the smallest reconstruction errors.	
	
\subsection{Skull-guided Face Re-synthesis (Inpainting)}
\label{Sec:SkullGuidedInpainting}
	
We compared our face inpainting algorithm with other strategies, including (1) FLAME-based 3DMM fitting (F3DMM, direct FLAME model fitting without considering skull geometry), (2) FLAME-based 3DMM fitting with geometric constraints from extended landmarks (F3DMM-GC), here we use the same regularization term $R = \alpha(\sum_{k=1}^{90}\alpha_k^2 + \sum_{k=1}^{90}\delta^2 + \sum_{k=1}^{15}\theta^2)$ which we used in Eqn.~\ref{Eqn:E_Loss} to do the fitting to avoid too large weight; (3) Generative Image Inpainting without geometric constraints~\cite{yeh2017semantic} (GII, direct inpainting using GAN, without considering skull geometry). Note that our algorithm is generative image inpainting with geometric constraints on extended landmarks. Hence, we denote it as GII-GC.
	
	\begin{figure*}[h]
		\centering
		\begin{tabular}{ccccccc}
			Skull & Candidate Image &  Broken face & F3DMM & F3DMM-GC & GII & GII-GC\\
			\includegraphics[height=0.14\textheight] {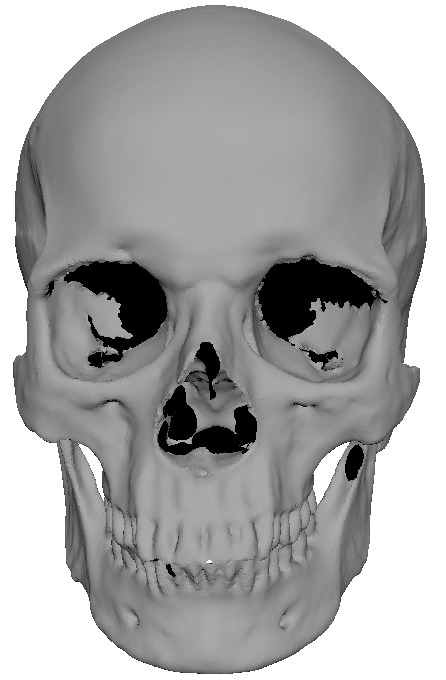} &
			\includegraphics[height=0.14\textheight] {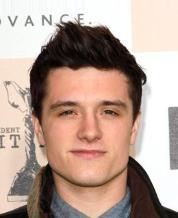} &
			\includegraphics[height=0.14\textheight] {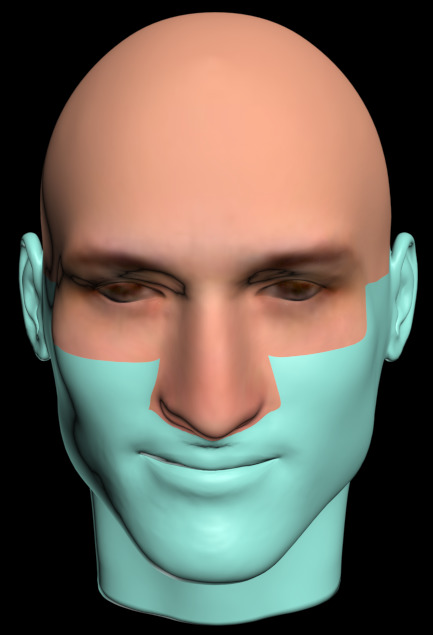} &
			\includegraphics[height=0.14\textheight] {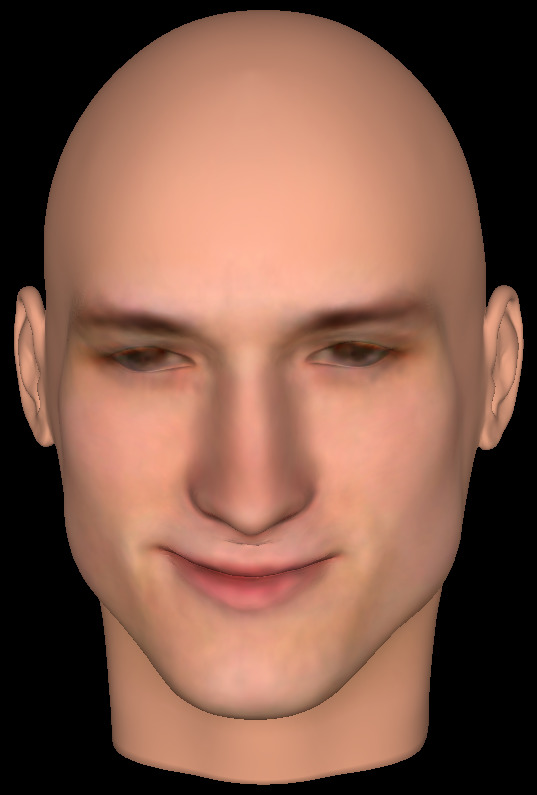} &
			\includegraphics[height=0.14\textheight] {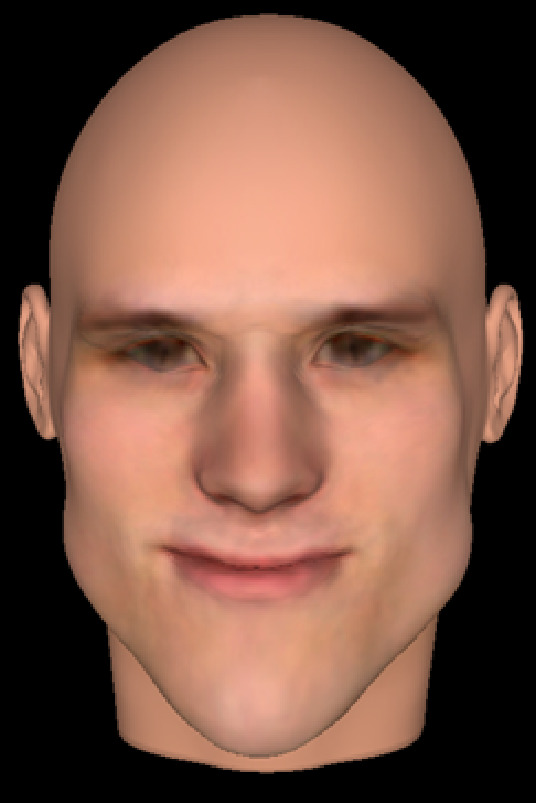} &
			\includegraphics[height=0.14\textheight] {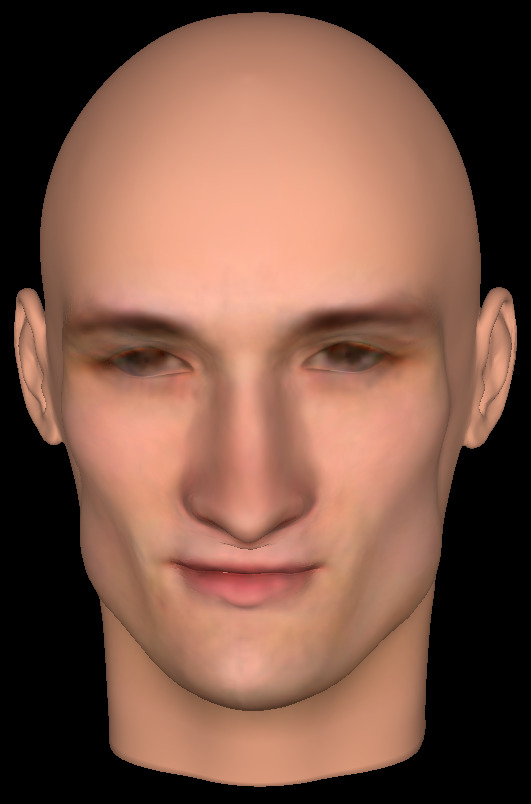} &
			\includegraphics[height=0.14\textheight] {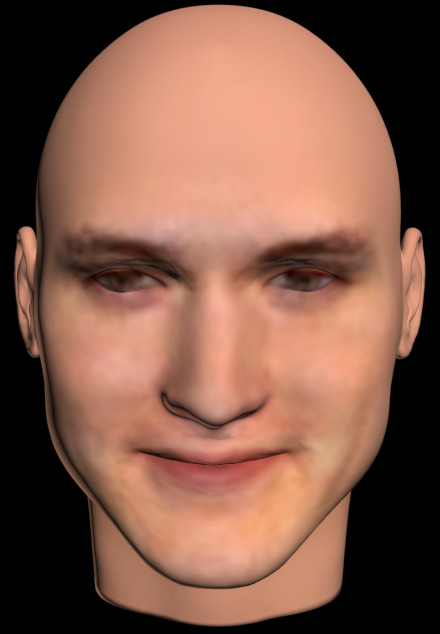} \\
			& 43.79\% & & 43.79\% &  100.0\% & 20.69\% &  100.0\% \\
			\includegraphics[height=0.14\textheight] {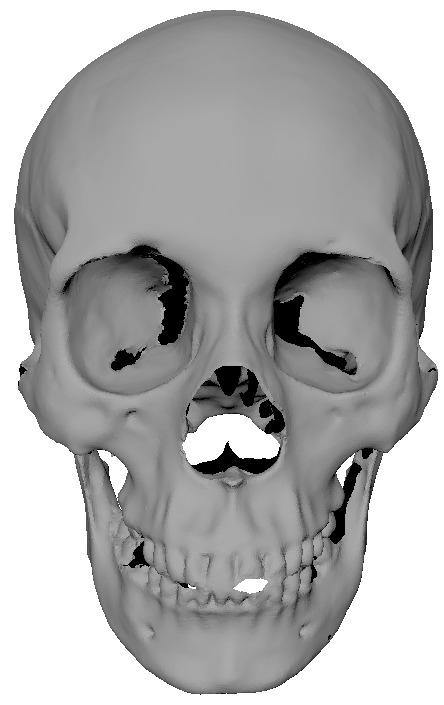} &
			\includegraphics[height=0.14\textheight] {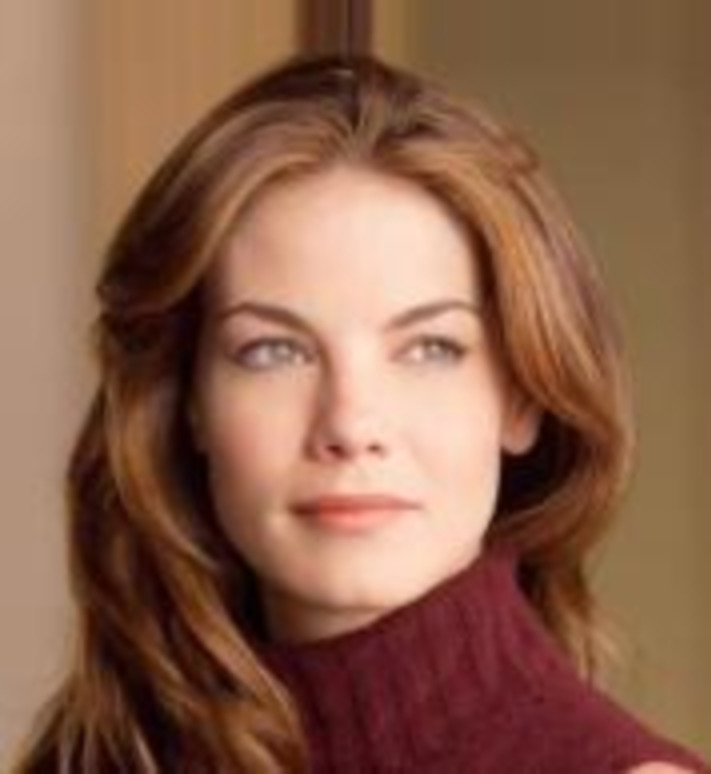} &
			\includegraphics[height=0.14\textheight] {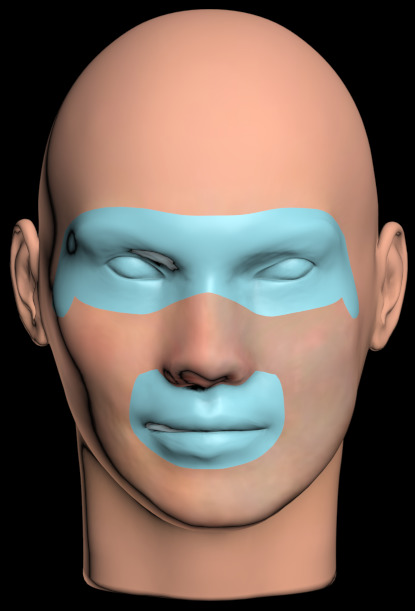} &
			\includegraphics[height=0.14\textheight] {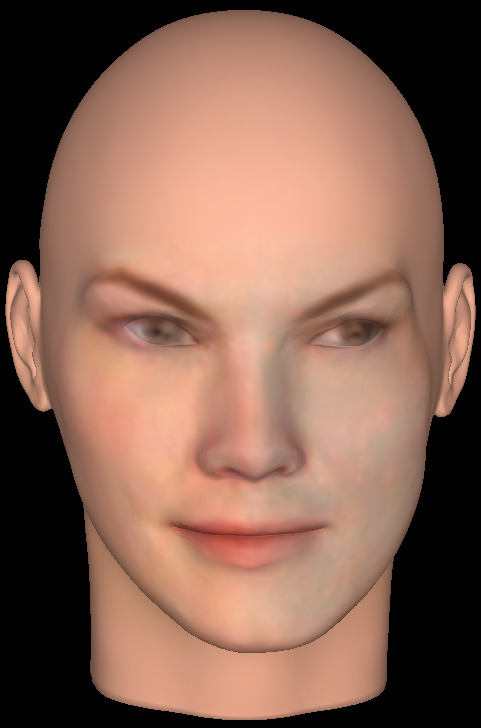} &
			\includegraphics[height=0.14\textheight] {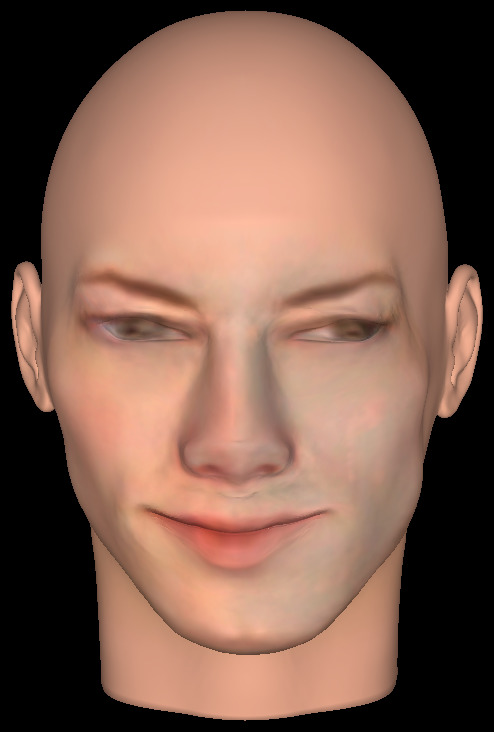} &
			\includegraphics[height=0.14\textheight] {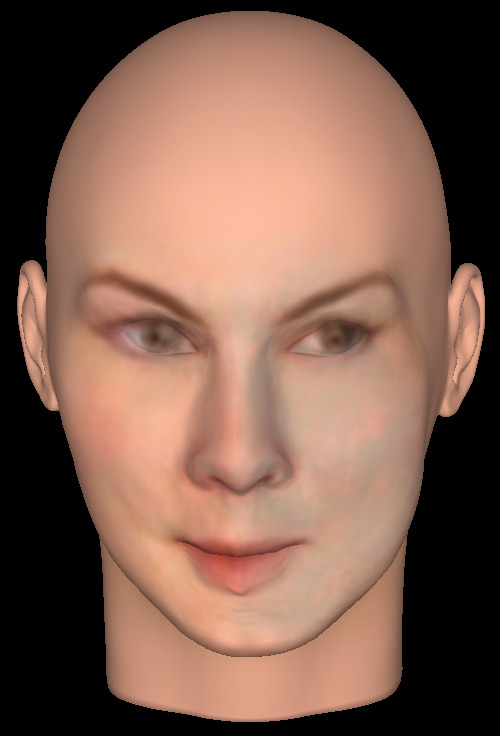} &
			\includegraphics[height=0.14\textheight] {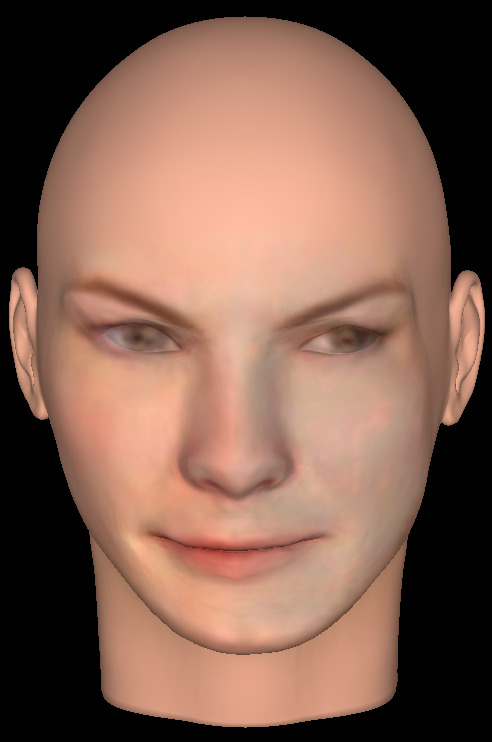} \\
			& 73.32\% & & 73.32\% &  100.0\% & 38.14\% &  100.0\% \\
			\includegraphics[height=0.14\textheight] {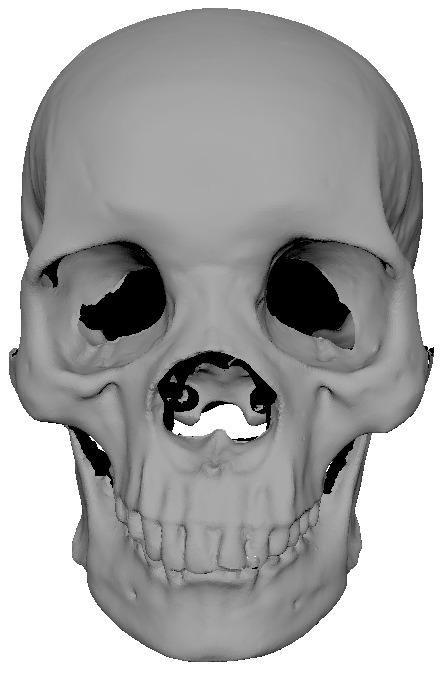} &
			\includegraphics[height=0.14\textheight] {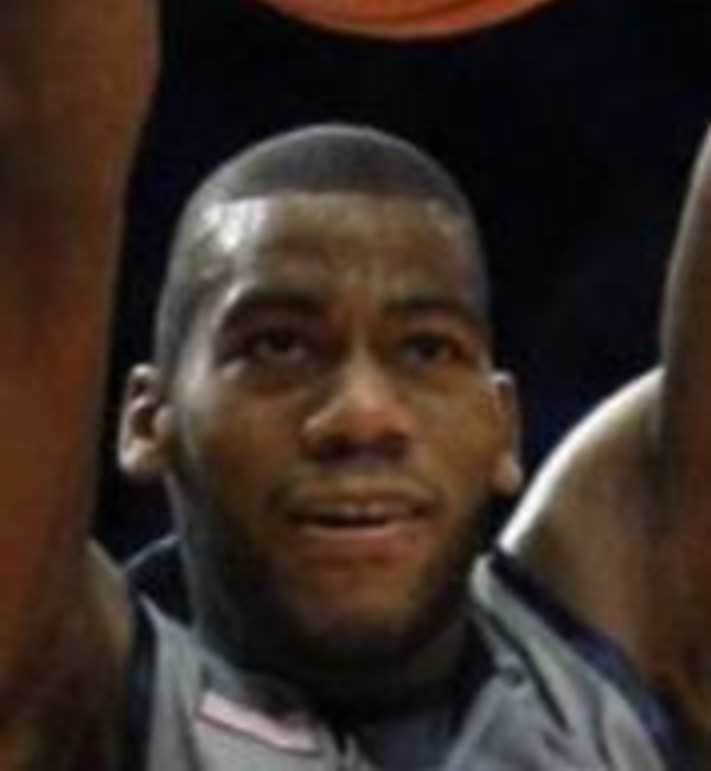} &
			\includegraphics[height=0.14\textheight] {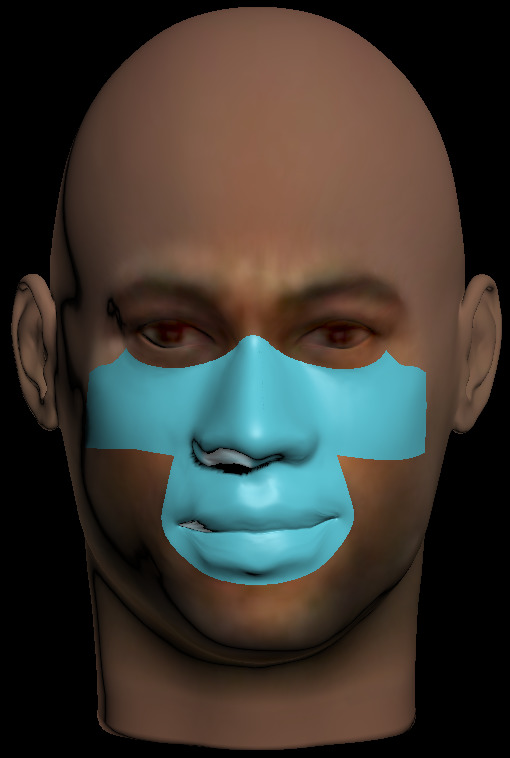} &
			\includegraphics[height=0.14\textheight] {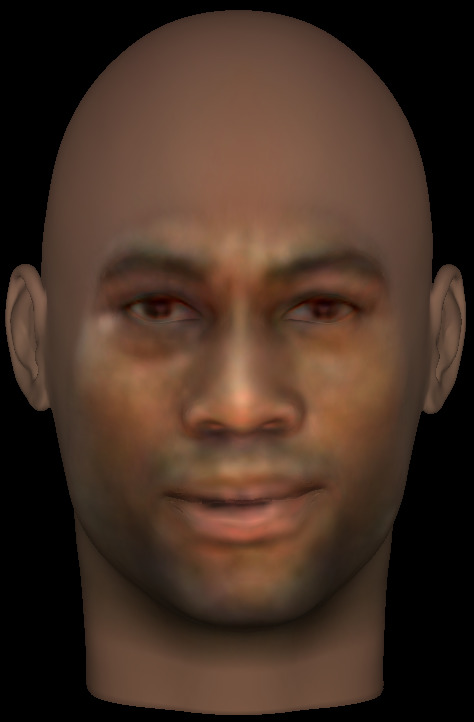} &
			\includegraphics[height=0.14\textheight] {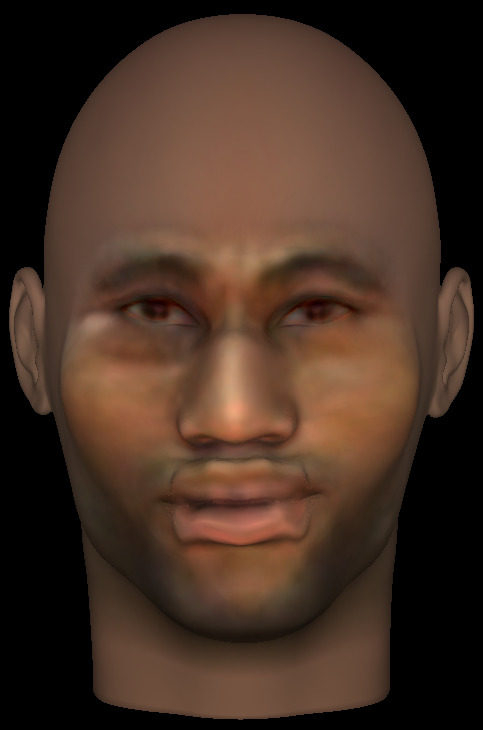} &
			\includegraphics[height=0.14\textheight] {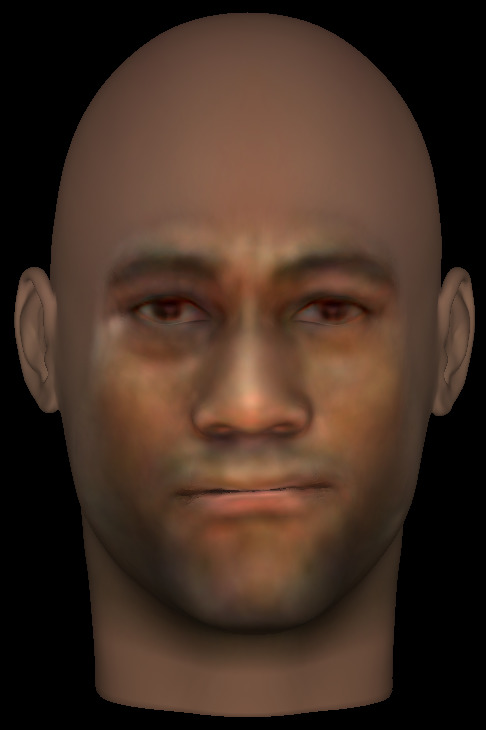} &
			\includegraphics[height=0.14\textheight] {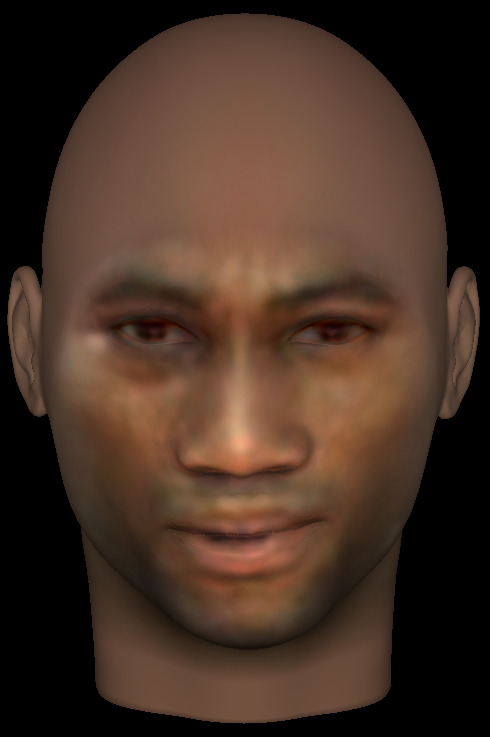} \\
			& 58.13\% & & 58.13\% &  100.0\% & 26.54\% &  100.0\% \\
			\includegraphics[height=0.14\textheight] {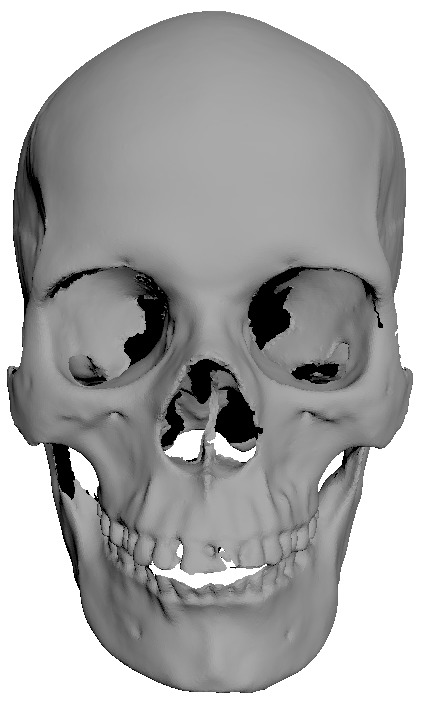} &
			\includegraphics[height=0.14\textheight] {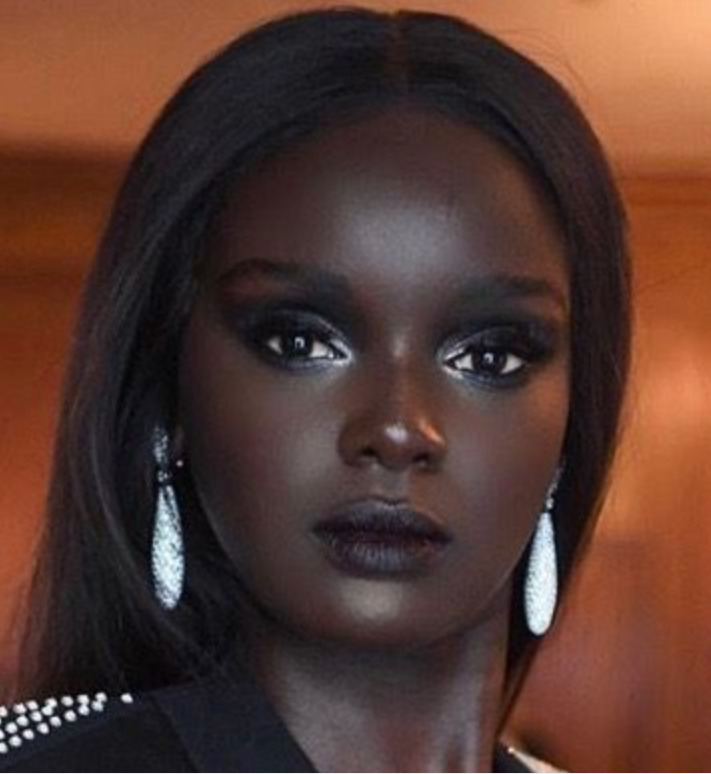} &
			\includegraphics[height=0.14\textheight] {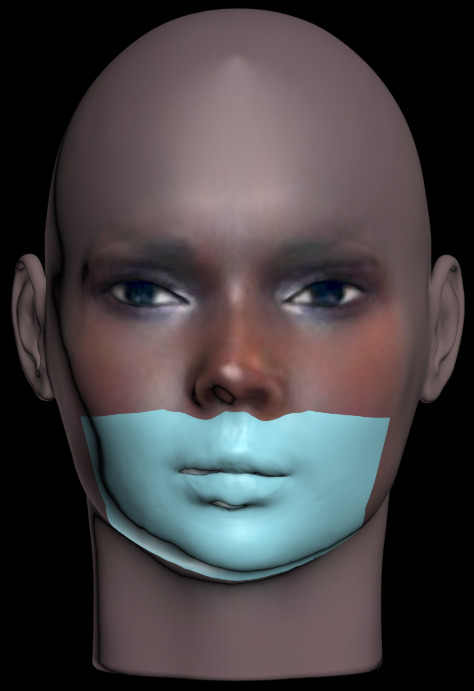} &
			\includegraphics[height=0.14\textheight] {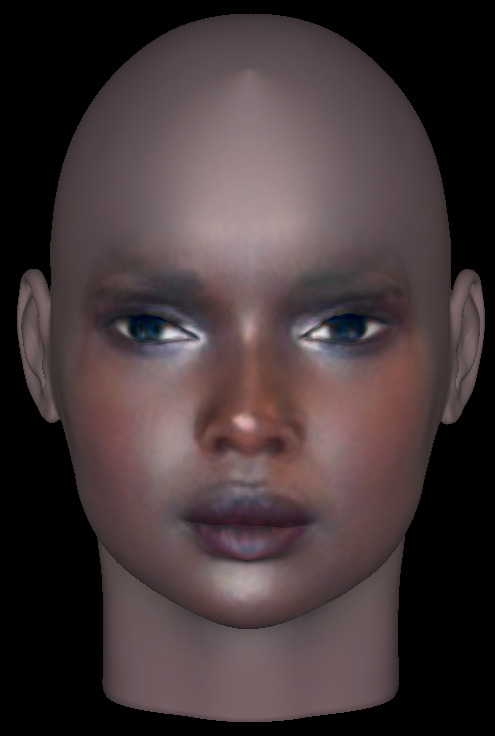} &
			\includegraphics[height=0.14\textheight] {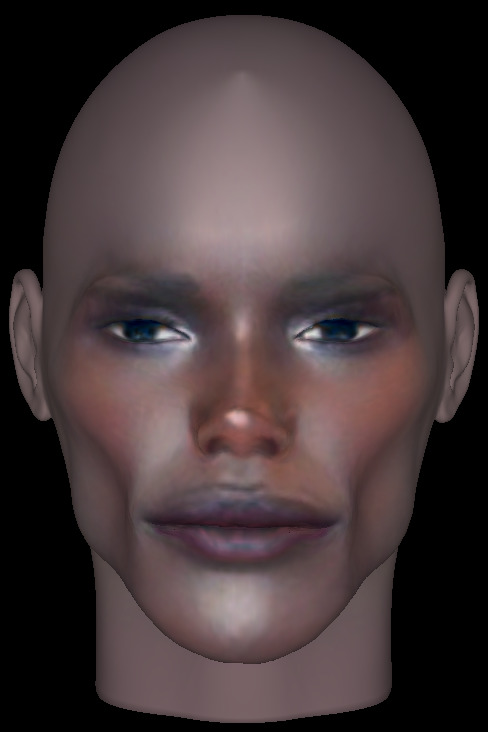} &
			\includegraphics[height=0.14\textheight] {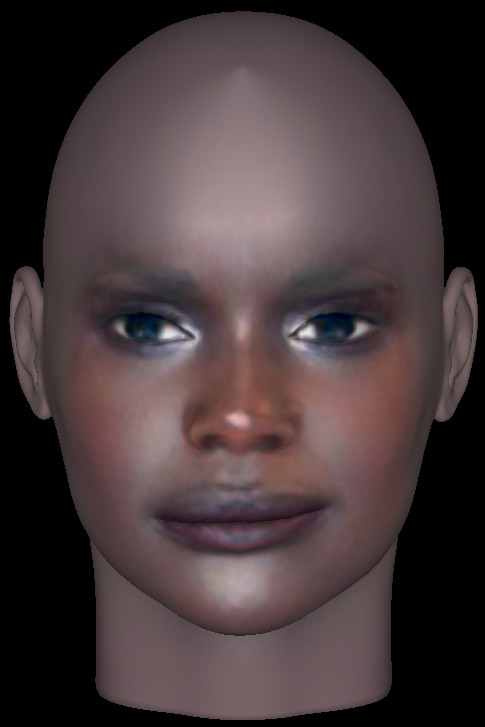} &
			\includegraphics[height=0.14\textheight] {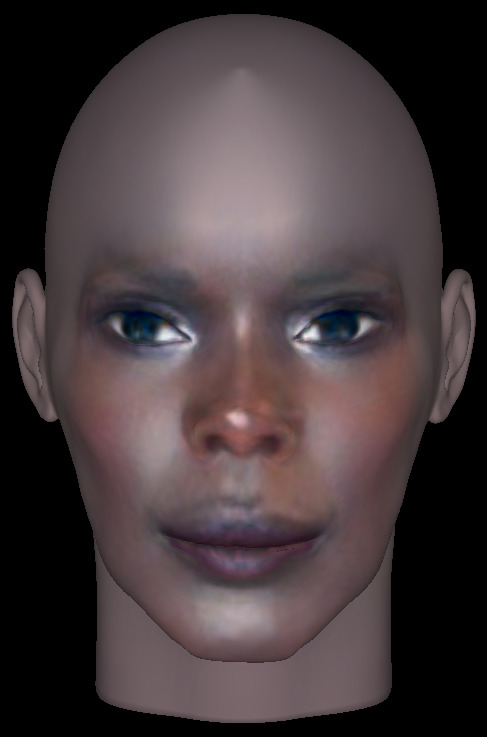} \\
			& 63.21\% & & 63.21\% &  100.0\% & 42.15\% &  100.0\% \\
		\end{tabular}
		\caption{Comparisons of Face Inpainting Results. Blue regions in broken face are low score regions which are removed before inpainting. The percentage values in sub-captions under Candidate Image is the inital superimposition score of this candidate. The percentage values in sub-captions under each algorithm are the final superimposition scores after inpainting. Indeed, direct inpainting without considering skull will not produce faces with high superimposition. scores.~\label{Fig:InpaintingComp}}
	\end{figure*}
	
	Fig.~\ref{Fig:InpaintingComp} illustrates two face inpainting results produced by different approaches. 	
	For a same query skull, the face inpainting on starts from two different initial faces, (a) and (f). The unmatched regions that need re-synthesis are colored in blue. 	
	The direct FLAME model fitting (F3DMM) just fills the holes and restores a face without considering the skull. Hence, it has low superimposition score. 
	The skull-guided modeling fitting (F3DMM-GC) produces well superimposed faces, but the face could be unrealistic and have artifacts especially near the constrained extended landmarks. 
	The generative image inpainting (GII) approach using our trained GAN will produce a realistic face, but also does not give us control on its alignment with the skull. 
	Using our geometrically constrained generative inpainting algorithm (GII-GC), we can obtain a realistic face that well matches with the skull. 
	From Fig.~\ref{Fig:InpaintingComp} (e,j), we can also see that the GII-GC inpainting is stable. The resultant faces are visually quite similar to each other.

	\subsection{Validation using CT Scans}
	
\begin{figure*}[h]
\centering
\begin{tabular}{cccccc}
\includegraphics[height=0.17\textwidth] {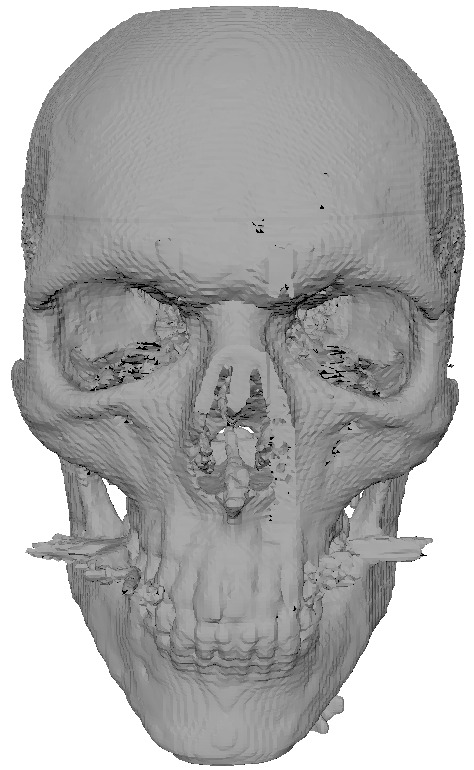} &
\includegraphics[height=0.17\textwidth] {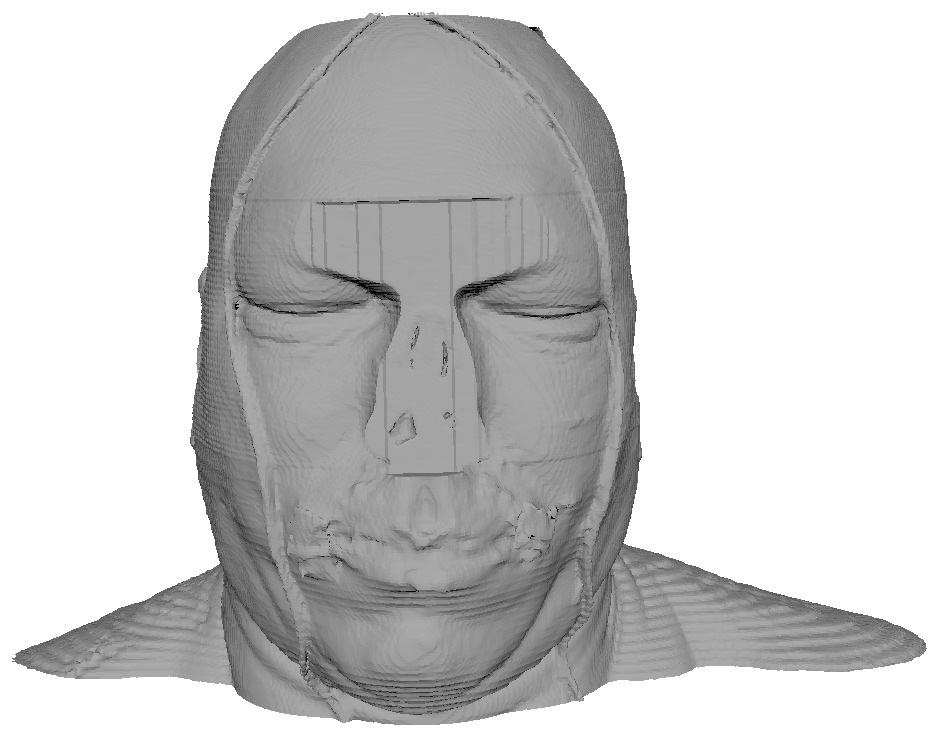} &
\includegraphics[height=0.17\textwidth] {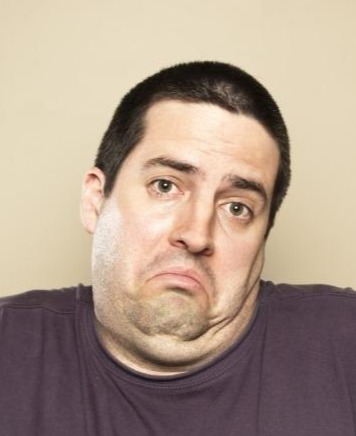} &
\includegraphics[height=0.17\textwidth] {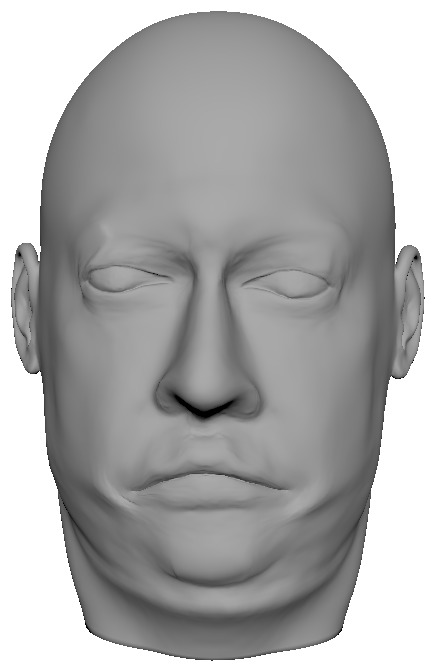}  &
\includegraphics[height=0.17\textwidth] {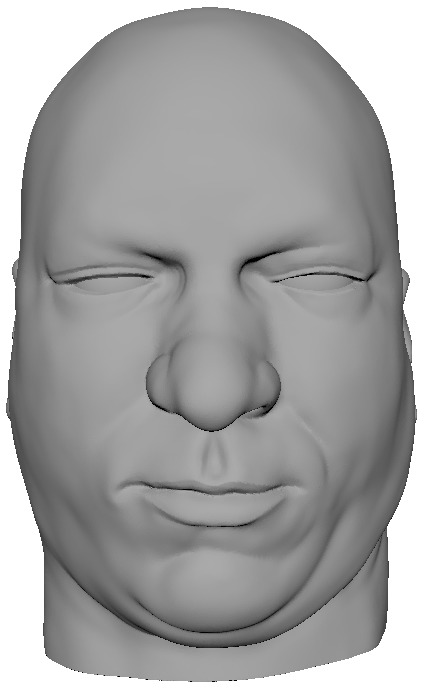}&
\includegraphics[height=0.17\textwidth] {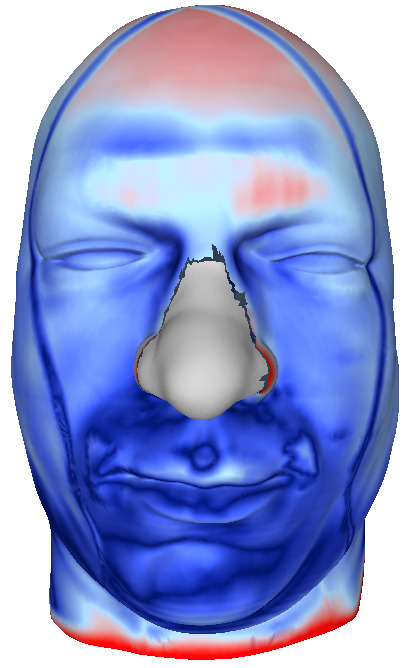} \\
(a) & (b) & (c) & (d) & (e) & (f) \\
\includegraphics[height=0.17\textwidth] {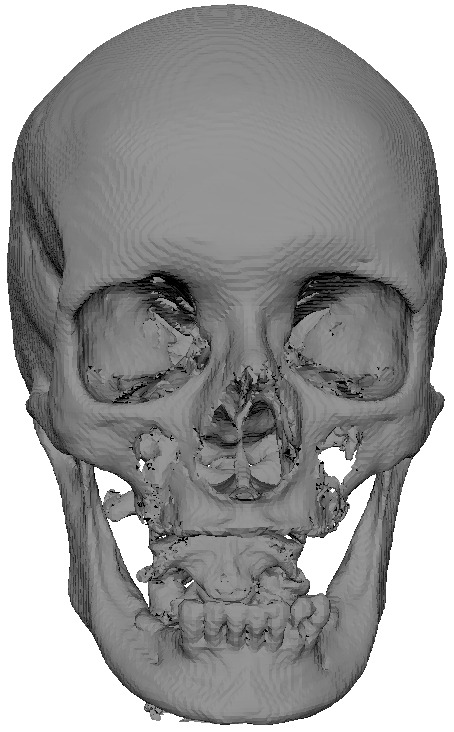} &
\includegraphics[height=0.17\textwidth] {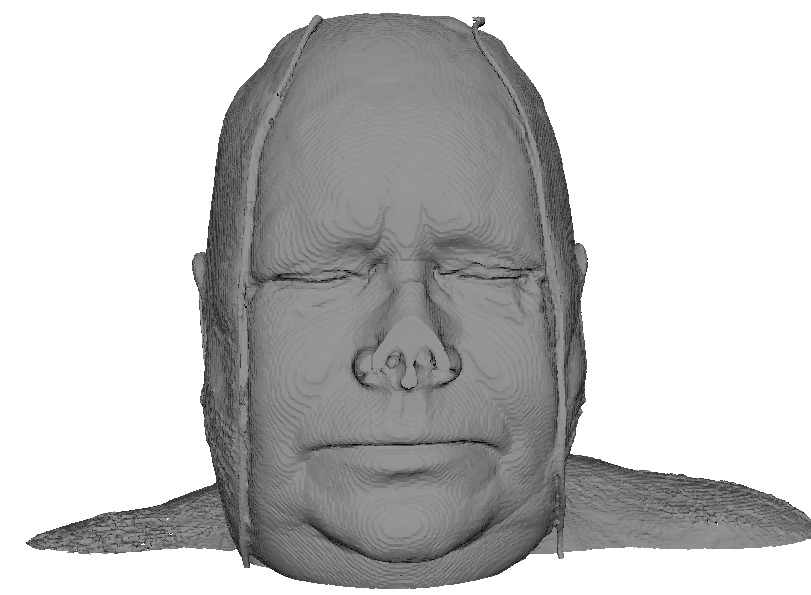} &
\includegraphics[height=0.17\textwidth] {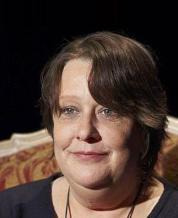} &
\includegraphics[height=0.17\textwidth] {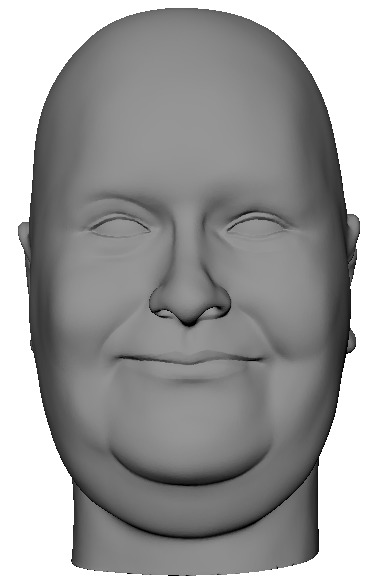}  &
\includegraphics[height=0.17\textwidth] {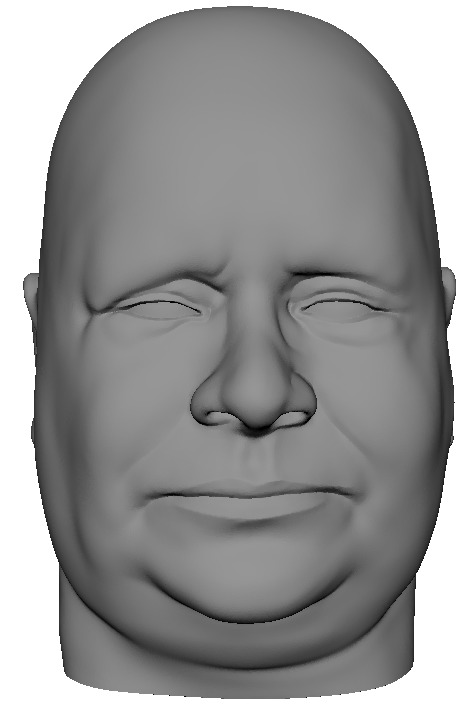} &
\includegraphics[height=0.17\textwidth] {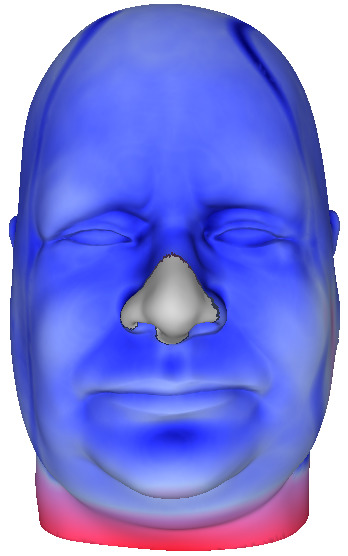}\\
(g) & (h) & (i) & (j) & (k) & (l) \\
\includegraphics[height=0.17\textwidth] {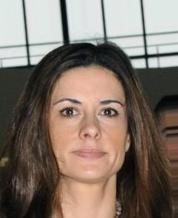} &
\includegraphics[height=0.17\textwidth] {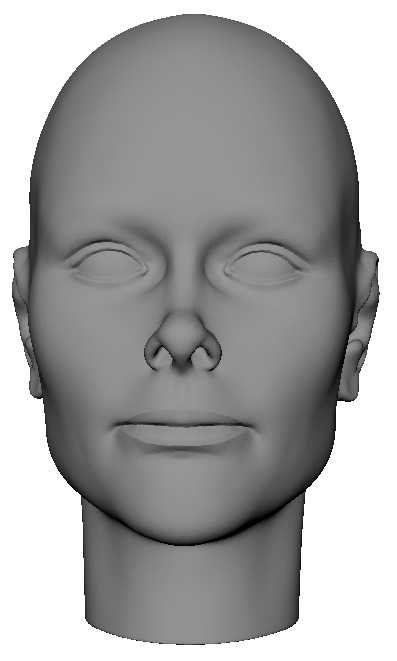} &
\includegraphics[height=0.17\textwidth] {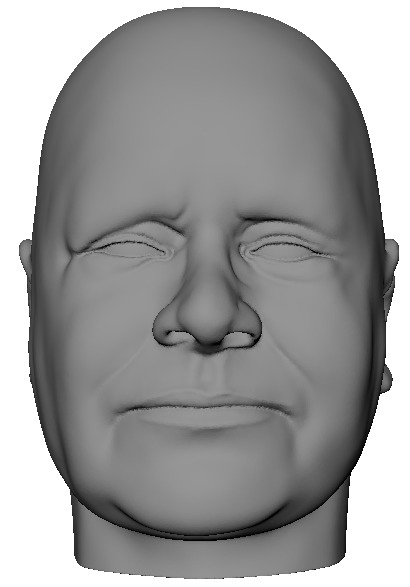} &
\includegraphics[height=0.17\textwidth] {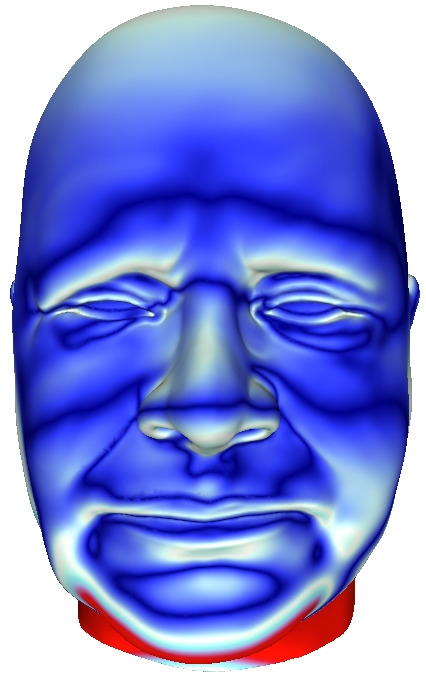}  &
\includegraphics[height=0.17\textwidth] {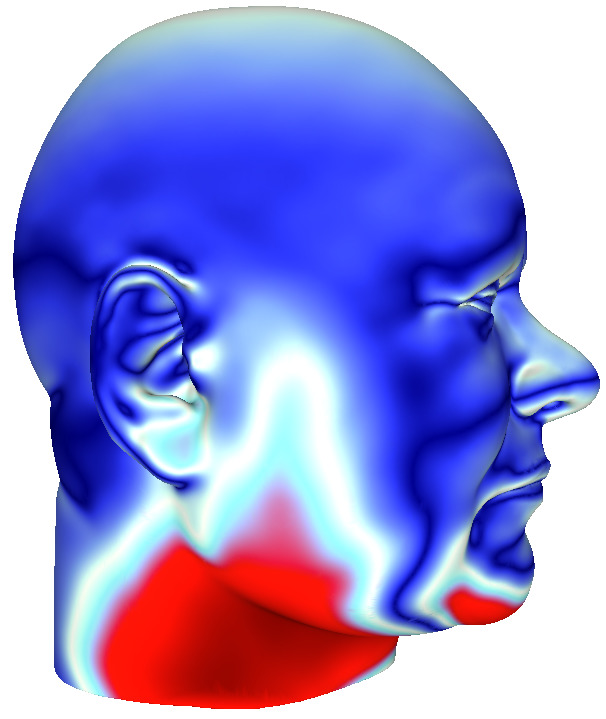} &
\includegraphics[height=0.17\textwidth] {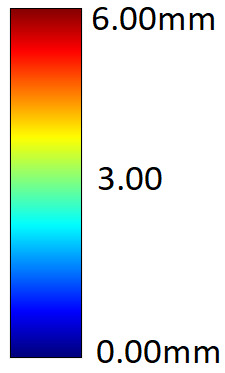}\\ 
(m) & (n) & (o) & (p) & (q) & (r) \\
\end{tabular}
\caption{Validation of Facial Reconstruction using CT Scans. 
(a, g) and (b, h) are skull and face iso-surfaces reconstructed from CT scans. 
(c, d) and (i, j) are the best matched face candidates, respectively. 
Their corresponding re-synthesized faces are shown in (e, k), and their deviation from the ground-truth (b, h) are color-encoded in (f, l) respectively. The average error in (f, l) are $2.7 mm$ and $2.3 mm$ respectively.
(m, n) are is another randomly selected candidate face for skull (g); accordingly, (o) is the final re-synthesized face. (p, q) show the color-encoded deviation from the ground-truth from two viewing angles. The colorbar is given in (r). \label{Fig:CTdiffStart}}
\end{figure*}

We collected two CT scans of human heads, and reconstructed the corresponding 3D skull and face geometries. 
Fig.~\ref{Fig:CTdiffStart} illustrates this experiment. 
(a) is the reconstructed query skull, and (b) is its corresponding face, serving as our face reconstruction ground-truth.
(c, d) is the found best-matched candidate face (image and 3D geometry, respectively). (e) is the final re-synthesized face. (f) is the color-encoded point-to-point error between (e) and the ground-truth (b). Note that the nose region is not available in (b), and hence it is colored in gray. 
Similarly, the second row illustrates the validation of a reconstruction conducted on another query skull.
The average error in (f) and (l) are $2.7 mm$ and $2.3 mm$ respectively, which are significantly 
smaller than the image-based face reconstruction error (see Table~\ref{Tab:ReconComp}).

To see how the candidate image affects the final reconstruction result. For skull (g) we use another random image (m) as the starting face (n). After the face is re-synthesized following the skull, the result is shown in (o). As the color-encoding in (p,q) shows, the frontal face region of (o) is very similar to (k) as this region is defined by many extended landmarks. The biggest difference is in the chin. Interestingly, the double-chin in (k) is from the candidate face (i) , while this does not exist in (o) as the thinner candidate (m) does not possess this characteristics.

\section{Conclusions}
\label{Sec:Conclusions}
We propose a new facial reconstruction pipeline for forensic skull identification. Unlike commonly adopted facial reconstruction approaches which directly reconstruct the face from the skull, we first generate many 3D face candidates from an image database. Then, we perform skull-face superimpositions to pick a best matched face candidate. Finally, we develop a new constrained generative model to modify the face to get a new face that matches well with the skull. 
To build an effective geometry constrained generative face inpainting model, we introduce a geometric loss term to better restrict the search space inside the latent space to obtain the globally realistic face that well matches the given skull. 
We demonstrate this pipeline can produce a stable facial reconstruction from a skull, starting from different face candidates.  
The accuracy of the reconstruction is also verified using skull-face pairs extracted from CT scans. 

\begin{figure*}[h!tb]
	\centering
	\begin{tabular}{ccc}
		\includegraphics[height=0.15\textheight] {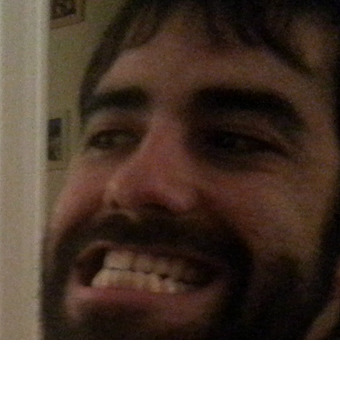} &
		\includegraphics[height=0.15\textheight] {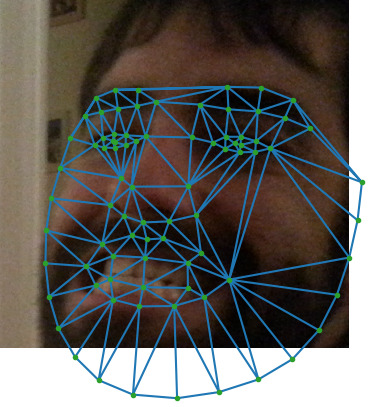} &
		\includegraphics[height=0.15\textheight] {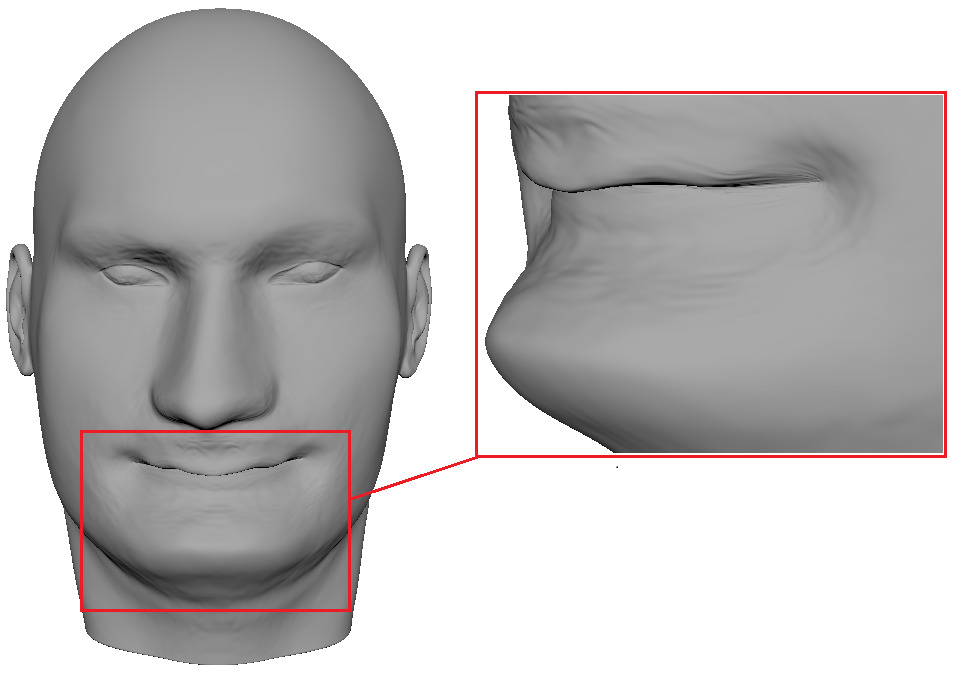} \\
		(a) & (b) & (c)\\
		\includegraphics[height=0.15\textheight] {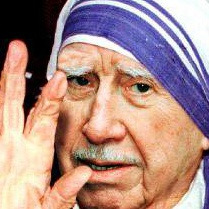} &
		\includegraphics[height=0.15\textheight] {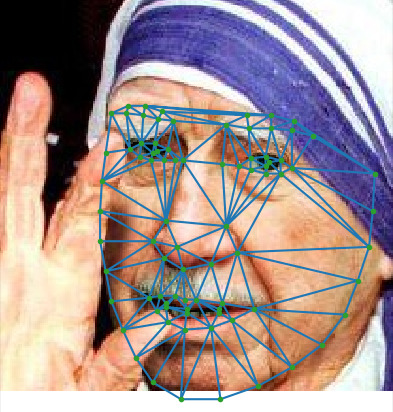} &
		\includegraphics[height=0.15\textheight] {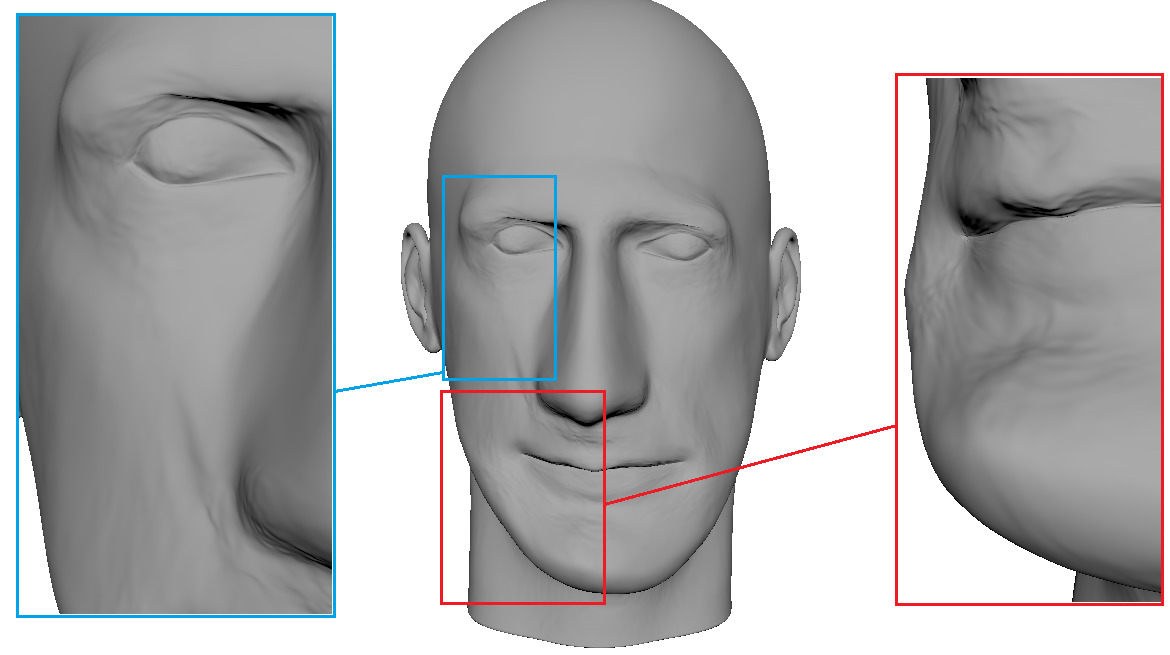} \\
		(d) & (e) & (f)\\
	\end{tabular}
	\caption{Failure examples. (a) A face image with bad viewing angle and exaggerated expression; (b) shows the detected landmarks; the reconstructed face (c) has unnatural geometry in the mouth and chin region. The face in (d) has some part blocked by his hand; its detected detected landmarks are shown in (e); the final reconstructed face (f) has artifact in left chin and right eight regions.~\label{Fig:FailureCases}}
\end{figure*}	

\textbf{Limitations and Future Work}. 
First, in \emph{image-based face reconstruction}, if the input photo has exaggerated expression, big occlusion, or is from a bad viewing angle, our encoder, whose loss relies on face landmark detection, may produce unrealistic reconstructions. 
Because on exaggerated expression, the detector may identify features incorrectly; and for photos with bad viewing angles, invisible landmarks are predicted and may also be inaccurate. 
Without accurate landmark identification, the reconstruction's performance would suffer. Fig. ~\ref{Fig:FailureCases} shows two failure cases. In (a), the camera angle is bad and the expression is far away from a calm face. Some out-of-image landmarks are estimated (b), which affect the face reconstruction result (c). 
In (c), we can see that geometry around the chin is unnatural, because most landmarks on this region are predicted (b) and are probably not very accurate. 
Also, the reconstructed faces from the FLAME model have closed mouths, and so when handling this widely opened mouth, the mouth geometry could also be inaccurate (and the direct texture). 
In (d), part of the face is blocked by a hand.
Some of the landmark around the left chin, one is out of the image and two are blocked by hands,  are predicted inaccurately (e). This leads to artifacts in the left cheek in the final reconstructed face (f).

Second, in skull-guided face re-synthesis, we convert the 3D geometric inpainting into image inpainting followed by a 3D reconstruction. 
A more natural design could be building a 3D face generator. However, the lack of large volume of 3D face datasets and effective deep geometric autoencoder architecture makes the construction of this 3D face generator challenging. 
We will explore possible direct 3D approaches in the near future. 
	
 \bibliographystyle{ACM-Reference-Format}
\bibliography{sig2018}
	
\end{document}